\documentclass[twocolumn,aps,showpacs,superscriptaddress,tightenlines,a4paper]{revtex4}

\usepackage{amssymb}
\usepackage{amsmath}
\usepackage{bbold}
\usepackage{mathrsfs}
\usepackage{paralist}
\usepackage{graphicx}
\usepackage{xcolor}

\begin{document}
%
%
	\title{Motion of macroscopic bodies in the electromagnetic field}
%
%
	\author{S. A. R. Horsley}
	\affiliation{Electromagnetic and Acoustic Materials Group, Department of Physics and Astronomy, University of Exeter, Stocker Road, Exeter, EX4 4QL, UK}
	\email{S.Horsley@exeter.ac.uk}
%
%
	\begin{abstract}
		A theory is presented for calculating the effect of the electromagnetic field on the centre of mass of a macroscopic dielectric body that is valid in both quantum and classical regimes.  We apply the theory to find the classical equation of motion for the centre of mass of a macroscopic object in a classical field, and the spreading of an initially localized wave--packet representing the centre of mass of a small object, in a quantum field.  The classical force is found to be consistent with the identification of the Abraham momentum with the mechanical momentum of light, and the motion of the wave-packet is found to be subject to an acceleration due to the Casimir force, and a time dependent fluctuating motion due the creation of pairs of excitations within the object.  The theory is valid for any dielectric that has susceptibilities satisfying the Kramers--Kr\"{o}nig relations, and is not subject to arguments regarding the form of the electromagnetic energy--momentum tensor within a medium.
	\end{abstract}
%
%
	\pacs{42.50.Wk,03.30.+p}
	\maketitle
	\bibliographystyle{unsrt}
%
%
	\par
	In recent years there has been significant progress in cooling macroscopic degrees of freedom down to temperatures where deviations from classical behaviour can be observed~\cite{marquardt2008,kippenberg2008,marquardt2009,aspelmeyer2012}.  This is to the extent that when sufficiently isolated from other degrees of freedom, the macroscopic mechanical motion can be reduced to the quantum mechanical ground state~\cite{teufel2011,naeini2011,chan2012}.  It thus becomes feasible to prepare superpositions of macroscopic states of matter~\cite{hammerer2009,romero2011}, where quantum mechanics is known to be at its most perplexing, and where the wave--function is particularly hard to take as a serious description of reality~\cite{wheeler1983,penrose2003}.
	\par
	Motivated by this remarkable work we look to contribute through asking how we should properly describe the motion of a macroscopic object interacting with a quantum field.  To be more specific, suppose that we have an object composed of \(\sim 10^{12}\) atoms, interacting with one another and the outside world through the electromagnetic field, and where the object may have been prepared in such a way that its centre of mass, \(\boldsymbol{R}\) can be considered as a quantum mechanical variable with an associated operator, \(\hat{\boldsymbol{R}}\) (see e.g.~\cite{chang2010,pender2012}).  We wish to understand the dynamics of the centre of mass in this situation.
	\par
	It is evident that we cannot make any progress if we work in terms of a microscopic description.  Yet, the comparatively macroscopic scale of the object has the advantage that we can assume that the collective internal atomic motion is close to some equilibrium configuration, and that any linear perturbation of the system away from this can be characterized via a susceptibility. We therefore look for a Hamiltonian that can be derived from an action principle, correctly describes the dynamics of the centre of mass and the electromagnetic field, and is formulated in terms of susceptibilities rather than atoms.
	\par
	This is certainly not the first attempt to tackle this sort of problem.  Some time ago C. K. Law worked out the theory of the quantized electromagnetic field in a cavity when one of the walls moves according to the rules of quantum mechanics~\cite{law1995}.  More recently there have been several papers investigating the quantum motion of the centre of mass of macroscopic dielectric objects within the electromagnetic field~\cite{cheung2011,cheung2012,pflanzer2012}, often treating the system in the context of the formalism of open quantum systems, and considering the interaction of the object with a single cavity mode.
	\par
	We emphasize that we do not \emph{yet} aim to find a good model for the cavity configuration, but rather look for a  theory that correctly treats the motion of macroscopic bodies in an electromagnetic field.  One major new ingredient we add to previous work is that we develop a description that fully includes the effects of dispersion and dissipation of the electromagnetic field within the body.  As a byproduct we shall find that we can also provide unambiguous expressions for classical forces on dielectric bodies, with results that may well be relevant to the continuing discussion of optical momentum in media~\cite{barnett2006,stallinga2006,pfeifer2007,she2008,padgett2008,barnett2010,barnett2010b,philbin2011b}.  In this context our work is perhaps most closely related to~\cite{stallinga2006}.
	\par
	One may alternatively wish to see the following work as a kind of generalization of Casimir--Lifshitz theory~\cite{volume9,philbin2011} to the situation where the centre of mass of the dielectric is treated as a dynamical variable.  Our findings may clarify the difference between the results of~\cite{raabe2005} and~\cite{pitaevskii2006,philbin2011}.
%
%
	\section{Action and classical motion\label{classical-section}}  
	\par
	We begin with the action necessary to describe the classical motion of a dielectric in an electromagnetic field.  The Lagrangian associated with the free non--relativistic motion of the centre of mass of a body must take the form, \(L_{\text{\tiny{CM}}}=\frac{1}{2}M(d\boldsymbol{R}/dt)^{2}\), where \(M\) is its mass, and \(d\boldsymbol{R}/dt\) is the centre of mass velocity.  Similarly the Lagrangian associated with the free electromagnetic field is also well known to be, \(\mathscr{L}_{\text{\tiny{F}}}=(\epsilon_{0}/2)(\boldsymbol{E}^{2}-c^{2}\boldsymbol{B}^{2})\)~\cite{volume2}.  The challenge is in how to characterize their interaction.
	\par
	The key to understanding the interaction between field and dielectric can be found in the work of Huttner and Barnett~\cite{huttner1992}, and the later more general findings of Philbin~\cite{philbin2010,philbin2012}.  They find that the dynamics of a system that is characterised by a linear susceptibility can be precisely mimicked using a field of simple harmonic oscillators.  To understand why this must be so, consider the polarization field of a stationary dielectric that responds linearly to the electric field,
	\begin{equation}
		\boldsymbol{P}(\boldsymbol{x},t)=\int_{-\infty}^{\infty}\frac{d\Omega}{2\pi}\chi_{\text{\tiny{EE}}}(\Omega)\tilde{\boldsymbol{E}}(\boldsymbol{x},\Omega)e^{-i\Omega t}\label{polarization}
	\end{equation}
	Due to the fact that \(\boldsymbol{P}\) is a function of the electric field in the past and not the future, \(\chi_{\text{\tiny{EE}}}(\Omega)\) must be an analytic function of frequency in the upper half of the complex plane~\cite{volume5,volume8}.  This has the consequence that the real and imaginary parts of \(\chi_{\text{\tiny{EE}}}\) are related by the Kramers--Kr\"onig relations
	\begin{equation}
		\text{Re}[\chi_{\text{\tiny{EE}}}(\Omega)]=\frac{2}{\pi}\text{P}\int_{0}^{\infty}\frac{\omega\text{Im}[\chi_{\text{\tiny{EE}}}(\omega)]}{(\omega^{2}-\Omega^{2})}d\omega\label{kramers-kronig}
	\end{equation}
	where `\(\text{P}\)' indicates the principal part of the integral.  Inserting (\ref{kramers-kronig}) into (\ref{polarization}) and using the convolution theorem, we find that the polarization field can be written as
	\begin{multline}
		\boldsymbol{P}(\boldsymbol{x},t)=\int_{0}^{\infty}d\omega\frac{2\omega\text{Im}[\chi_{\text{\tiny{EE}}}(\omega)]}{\pi}\\
		\times\int_{-\infty}^{t}dt^{\prime}G_{\omega}(t-t^{\prime})\boldsymbol{E}(\boldsymbol{x},t^{\prime})\label{polarization-2}
	\end{multline}
	\par
	Within (\ref{polarization-2}) we have the quantity, \(G_{\omega}(t-t^{\prime})=\Theta(t-t^{\prime})\sin[\omega(t-t^{\prime})]/\omega\), which is recognised as the retarded Green function for a simple harmonic oscillator of unit mass and natural frequency \(\omega\)~\cite{volume1,stallinga2006}.  Expression (\ref{polarization-2}) could also be obtained if the polarization field were identified with an integral over a continuum of simple harmonic oscillators, all coupled linearly to the electric field, each oscillator having a mass \((2\omega\text{Im}[\chi_{\text{\tiny{EE}}}(\omega)]/\pi)^{-1}\).
	\par
	With this in mind, the Lagrangian density describing a \emph{stationary} dielectric in an electromagnetic field is
	\begin{multline}
		\mathscr{L}=\mathscr{L}_{\text{\tiny{F}}}+\boldsymbol{P}\boldsymbol{\cdot}\boldsymbol{E}+\boldsymbol{M}\boldsymbol{\cdot}\boldsymbol{B}\\
		+\frac{1}{2}\sum_{\lambda}\int_{0}^{\infty}d\omega\left[\dot{\boldsymbol{X}}_{\omega}^{(\lambda)\,2}-\omega^{2}\boldsymbol{X}_{\omega}^{(\lambda)\,2}\right],\label{stationary-lagrangian}
	\end{multline}
	where \(\lambda\) takes two indices, \emph{E} and \emph{B}, these labelling the two sets of oscillators~\cite{philbin2010}.  The polarization and magnetization are given by
	 \begin{align}
	 	\boldsymbol{P}=\int_{0}^{\infty}\alpha(\omega)\boldsymbol{X}_{\omega}^{(\text{\tiny{E}})}d\omega\nonumber\\
		\boldsymbol{M}=\int_{0}^{\infty}\beta(\omega)\boldsymbol{X}_{\omega}^{(\text{\tiny{B}})}d\omega\label{polmag}
	\end{align}
	where \(\alpha(\omega)=\sqrt{2\omega\text{Im}[\chi_{\text{\tiny{EE}}}(\omega)]/\pi}\) and \(\beta(\omega)=\sqrt{2\omega\text{Im}[\chi_{\text{\tiny{BB}}}(\omega)]/\pi}\).  Note the equivalence between the imaginary parts of the susceptibilities appearing within the mass of the oscillator field (as in the previous paragraph), and appearing within the coupling, as in (\ref{polmag}).  From this action a Hamiltonian can be derived, and the system can be canonically quantized.  Recent work has found (\ref{stationary-lagrangian}) to provide a rigorous basis for the theory of the Casimir effect~\cite{philbin2010}
	  \par
	  When the dielectric is set in motion, the polarization and magnetization are no longer given by (\ref{polmag}), but to first order in \(\dot{\boldsymbol{R}}\) transform as, \(\boldsymbol{P}^{\prime}=\boldsymbol{P}+\dot{\boldsymbol{R}}\boldsymbol{\times}\boldsymbol{M}/c^{2}\), and \(\boldsymbol{M}^{\prime}=\boldsymbol{M}-\dot{\boldsymbol{R}}\boldsymbol{\times}\boldsymbol{P}\).  Furthermore, the susceptibility becomes non--local, and a function of the Doppler shifted frequency, \(\Omega^{\prime}=\Omega-\dot{\boldsymbol{R}}\boldsymbol{\cdot}\boldsymbol{k}\)~\cite{volume8}.  Taking these factors into account, the action for a uniformly moving dielectric was derived in~\cite{horsley2011b} and from this a quantum theory moving media was constructed~\cite{horsley2012}.  The action was found to be
	\begin{equation}
		S[A^{\mu},\boldsymbol{X}_{\omega}^{(\lambda)}]=\int \left[\mathscr{L}_{\text{\tiny{F}}}+\mathscr{L}_{\text{\tiny{INT}}}+\mathscr{L}_{\text{\tiny{R}}}\right]d^{4}x\label{initial-action}
	\end{equation}
	where we write the electromagnetic field in terms of the four--vector potential, \(\boldsymbol{E}=-\boldsymbol{\nabla}\varphi-\dot{\boldsymbol{A}}\), \(\boldsymbol{B}=\boldsymbol{\nabla}\boldsymbol{\times}\boldsymbol{A}\).  The Lagrangian density associated with the two sets of simple harmonic oscillators is modified from the stationary case,
	\begin{multline}
		\mathscr{L}_{\text{\tiny{R}}}=\frac{1}{2}\sum_{\lambda}\int_{0}^{\infty}\bigg\{\left[\left(\frac{\partial}{\partial t}+\dot{\boldsymbol{R}}\boldsymbol{\cdot}\boldsymbol{\nabla}\right)\boldsymbol{X}^{(\lambda)}_{\omega}\right]^{2}\\
		-\omega^{2}{\boldsymbol{X}^{(\lambda)}_{\omega}}^{2}\bigg\}d\omega
	\end{multline}
	and the interaction Lagrangian density is also changed, due to the aforementioned transformation of the polarization and the magnetization,
	\begin{equation}
		\mathscr{L}_{\text{\tiny{INT}}}=\sum_{\lambda}\int_{0}^{\infty}d\omega\left[\boldsymbol{E}\boldsymbol{\cdot}\boldsymbol{\alpha}_{\text{\tiny{E$\lambda$}}}(\omega)+\boldsymbol{B}\boldsymbol{\cdot}\boldsymbol{\alpha}_{\text{\tiny{B$\lambda$}}}(\omega)\right]\boldsymbol{\cdot}\boldsymbol{X}_{\omega}^{(\lambda)}\label{l_int}
	\end{equation}
	The \(\boldsymbol{\alpha}_{\text{\tiny{$\lambda$$\lambda^{\prime}$}}}\) matrices being given by, \(\boldsymbol{\alpha}_{\text{\tiny{EE}}}=\boldsymbol{\mathbb{1}}_{3}\alpha(\omega)\); \(\boldsymbol{\alpha}_{\text{\tiny{BB}}}=\boldsymbol{\mathbb{1}}_{3}\beta(\omega)\); \(\boldsymbol{\alpha}_{\text{\tiny{EB}}}=(1/c^{2})\dot{\boldsymbol{R}}\boldsymbol{\times}\boldsymbol{\mathbb{1}}_{3}\beta(\omega)\); and \(\boldsymbol{\alpha}_{\text{\tiny{BE}}}=-\dot{\boldsymbol{R}}\boldsymbol{\times}\boldsymbol{\mathbb{1}}_{3}\alpha(\omega)\).  In the above expressions we have suppressed the spatial dependence of the susceptibilities, which depend on the difference between the integration variable, \(\boldsymbol{x}\) and the position of the centre of mass, \(\boldsymbol{R}\), i.e. \(\chi_{\text{\tiny{EE}}}=\chi_{\text{\tiny{EE}}}(\boldsymbol{x}-\boldsymbol{R},\omega)\), being zero everywhere outside the dielectric.
	\par
	A comparison of expressions (\ref{initial-action}--\ref{l_int}) and those of~\cite{horsley2011b,horsley2012} will show that we are working under the approximation that \((\dot{\boldsymbol{R}}/c)^{2}\sim0\).  This is because we will now treat the motion of the centre of mass as a dynamical variable rather than an external parameter.  Beyond first order in \(\dot{\boldsymbol{R}}/c\) the dynamics of the centre of mass of an object are extremely subtle.  Properly speaking the centre of mass becomes a redundant variable, and one must instead deal with the centre of energy~\cite{volume2,shockley1967,coleman1968,aharonov1988,horsley2006}.
	\par
	As a final comment we note that we only consider the translational motion of the body, and not its rotation about the centre.  It is possible to include this through introducing the three Euler angles into spatial dependence of the quantities, \(\boldsymbol{\alpha}_{\lambda\lambda^{\prime}}\), but we do not study these effects here.

%
%
	\subsection{Electromagnetic force on a dielectric in free space\label{classical-radiation-pressure}}
	\par
	The action given in (\ref{initial-action}) allows us to calculate an unambiguous expression for the force on the centre of mass of a dielectric body.  To do this we need only add the kinetic energy term, \(L_{\text{\tiny{CM}}}\) to the Lagrangian,
	\begin{equation}
		S^{\prime}=\int L^{\prime} dt= S+\frac{1}{2}\int M\dot{\boldsymbol{R}}^{2} dt\label{modified-action}
	\end{equation}
	Then the classical equations of motion of the centre of mass can be obtained from an application of the Euler--Lagrange equations to the variable, \(\boldsymbol{R}(t)\)
	\begin{equation}
		\frac{d}{dt}\left(\frac{\partial L^{\prime}}{\partial \dot{\boldsymbol{R}}}\right)=\frac{\partial L^{\prime}}{\partial\boldsymbol{R}}\label{lagrange-eqn}
	\end{equation}
	We find that the associated canonical momentum is given by
	\begin{multline}
		\boldsymbol{p}=\frac{\partial L^{\prime}}{\partial\dot{\boldsymbol{R}}}=\boldsymbol{\mu}\boldsymbol{\cdot}\dot{\boldsymbol{R}}+\int\bigg(\frac{1}{c^{2}}\boldsymbol{M}\boldsymbol{\times}\boldsymbol{E}-\boldsymbol{P}\boldsymbol{\times}\boldsymbol{B}\bigg)d^{3}\boldsymbol{x}\\
		+\sum_{\lambda}\int d^{3}\boldsymbol{x}\int_{0}^{\infty}d\omega(\boldsymbol{\nabla}\boldsymbol{\otimes}\boldsymbol{X}_{\omega}^{(\lambda)})\boldsymbol{\cdot}\dot{\boldsymbol{X}}_{\omega}^{(\lambda)}\label{can-mom}
	\end{multline}
	which includes a contribution from the both the harmonic oscillators alone, and the coupling of the electromagnetic field to the oscillators.  The polarization and magnetization are the rest frame expressions given in (\ref{polmag}), and the tensorial quantity playing the role of the mass, \(\boldsymbol{\mu}\) contains contributions from both the intertial mass of the dielectric and the spatial distribution of the oscillator amplitudes,
	\begin{equation}
	\boldsymbol{\mu}=M\boldsymbol{\mathbb{1}}_{3}+\sum_{\lambda}\int d^{3}\boldsymbol{x}\int_{0}^{\infty}d\omega(\boldsymbol{\nabla}\boldsymbol{\otimes}\boldsymbol{X}_{\omega}^{(\lambda)})\boldsymbol{\cdot}(\boldsymbol{X}_{\omega}^{(\lambda)}\boldsymbol{\otimes}\overleftarrow{\boldsymbol{\nabla}})
	\end{equation}
	The second term in (\ref{can-mom})---involving a coupling of the oscillators and the fields---was also found in~\cite{horsley2006}.  However, the second line of (\ref{can-mom}) and second term within \(\boldsymbol{\mu}\) appear to be new.  The origin of these terms can be traced back to the continuum description of the medium, and ultimately the Doppler shift within the Kramer--Kr\"onig relations.  For example such a contribution is not present within the single oscillator model of the internal degrees of freedom given in~\cite{horsley2008}.
	\par
	To calculate the right hand side of (\ref{lagrange-eqn}) we note that the Lagrangian depends on \(\boldsymbol{R}\) only through the coupling matrices, \(\boldsymbol{\alpha}_{\lambda\lambda^{\prime}}\) within \(L_{\text{\tiny{INT}}}=\int \mathscr{L}_{\text{\tiny{INT}}}d^{3}\boldsymbol{x}\).  The derivative of this quantity with respect to the \(i\)th component of the centre of mass vector is found to be
	\begin{equation}
		\frac{\partial L^{\prime}}{\partial R_{i}}=\int d^{3}\boldsymbol{x}\sum_{\lambda}\int_{0}^{\infty}d\omega\left[\boldsymbol{E}\boldsymbol{\cdot}\frac{\partial\boldsymbol{\alpha}_{\text{\tiny{E$\lambda$}}}}{\partial R_{i}}+\boldsymbol{B}\boldsymbol{\cdot}\frac{\partial\boldsymbol{\alpha}_{\text{\tiny{B$\lambda$}}}}{\partial R_{i}}\right]\boldsymbol{\cdot}\boldsymbol{X}_{\omega}^{(\lambda)}\label{pot-grad}
	\end{equation}
	Due to the fact that the coupling only depends on the difference between the integration variable, \(\boldsymbol{x}\) and the centre of mass position we can equivalently write, \(\partial\boldsymbol{\alpha}_{\lambda\lambda^{\prime}}/\partial R_{i}=-\partial\boldsymbol{\alpha}_{\lambda\lambda^{\prime}}/\partial x_{i}\).  Inserting (\ref{can-mom}) and (\ref{pot-grad}) into (\ref{lagrange-eqn}) then gives an expression for the electromagnetic force on a moving dielectric body,
	\begin{widetext}
	\begin{multline}
		M\ddot{\boldsymbol{R}}+\left(\frac{\partial}{\partial t}+\boldsymbol{\dot{R}}\boldsymbol{\cdot}\boldsymbol{\nabla}_{\boldsymbol{R}}\right)\int\bigg(\frac{1}{c^{2}}\boldsymbol{M}\boldsymbol{\times}\boldsymbol{E}-\boldsymbol{P}\boldsymbol{\times}\boldsymbol{B}\bigg)d^{3}\boldsymbol{x}\\
		=-\int d^{3}\boldsymbol{x}\left[\boldsymbol{\nabla}\boldsymbol{\otimes}\left(\boldsymbol{P}+\frac{1}{c^{2}}\dot{\boldsymbol{R}}\boldsymbol{\times}\boldsymbol{M}\right)\boldsymbol{\cdot}\boldsymbol{E}+\boldsymbol{\nabla}\boldsymbol{\otimes}\left(\boldsymbol{M}-\dot{\boldsymbol{R}}\boldsymbol{\times}\boldsymbol{P}\right)\boldsymbol{\cdot}\boldsymbol{B}\right]\label{result-1}
	\end{multline}
	\end{widetext}
	To obtain (\ref{result-1}), which does not make explicit reference to the oscillator fields, we integrated out a term equal to the gradient of a scalar containing only oscillator amplitudes and not the electromagnetic field.  In the region outside the body where the \(\boldsymbol{\alpha}_{\lambda\lambda^{\prime}}=0\), the oscillator field is assumed to be at rest, therefore the divergence of any quantity containing them will integrate to zero.  We also applied the equations of motion for the two sets of oscillators,
	\begin{equation}
		\left[\left(\frac{\partial}{\partial t}+\boldsymbol{\dot{R}}\boldsymbol{\cdot}\boldsymbol{\nabla}\right)^{2}+\omega^{2}\right]\boldsymbol{X}_{\omega}^{(\lambda)}=\boldsymbol{\alpha}_{\text{\tiny{E$\lambda$}}}^{T}\boldsymbol{\cdot}\boldsymbol{E}+\boldsymbol{\alpha}_{\text{\tiny{B$\lambda$}}}^{T}\boldsymbol{\cdot}\boldsymbol{B}\label{osc-eqn}
	\end{equation}
	To further simplify (\ref{result-1}) we need to impose the equations of motion for the electromagnetic field.  These follow from a variation of the action (\ref{initial-action}) with respect to the four--potential, \(A^{\mu}=(\varphi/c,\boldsymbol{A})\), and are the familiar expressions; \(\boldsymbol{\nabla}\boldsymbol{\cdot}\boldsymbol{D}=0\); \(\boldsymbol{\nabla}\boldsymbol{\cdot}\boldsymbol{B}=0\); \(\boldsymbol{\nabla}\boldsymbol{\times}\boldsymbol{E}=-\partial\boldsymbol{B}/\partial t\); and \(\boldsymbol{\nabla}\boldsymbol{\times}\boldsymbol{H}=\dot{\boldsymbol{D}}\).  The auxiliary fields being given by \(\boldsymbol{D}=\epsilon_{0}\boldsymbol{E}+\boldsymbol{P}+(1/c^{2})\dot{\boldsymbol{R}}\boldsymbol{\times}\boldsymbol{M}\) and \(\boldsymbol{H}=\mu_{0}^{-1}\boldsymbol{B}-\boldsymbol{M}+\dot{\boldsymbol{R}}\boldsymbol{\times}\boldsymbol{P}\)~\footnote{Note that these definitions of \(\boldsymbol{D}\) and \(\boldsymbol{H}\) differ from those used when the polarization or magnetization are non--zero in the absence of an electromagnetic field.  Classically this is the case when a body has a permanent polarization or magnetization.  When quantum mechanical effects are taken into account the situation is more subtle for it is not possible for the electromagnetic field to be zero, so neither is it possible for the polarization and magnetization to vanish~\cite{philbin2010}.}.  Applying these equations of motion to (\ref{result-1}) gives,
	\begin{widetext}
	\begin{multline}
		M\ddot{\boldsymbol{R}}+\frac{1}{c^{2}}\frac{\partial}{\partial t}\int_{V}\left[\boldsymbol{E}\boldsymbol{\times}\boldsymbol{H}-\boldsymbol{E}\boldsymbol{\times}\left(\dot{\boldsymbol{R}}\boldsymbol{\times}\boldsymbol{P}\right)-\boldsymbol{B}\boldsymbol{\times}\left(\dot{\boldsymbol{R}}\boldsymbol{\times}\boldsymbol{M}\right)\right]d^{3}\boldsymbol{x}\\
		+\dot{\boldsymbol{R}}\boldsymbol{\cdot}\boldsymbol{\nabla}_{\boldsymbol{R}}\int_{V}\left(\frac{1}{c^{2}}\boldsymbol{M}\boldsymbol{\times}\boldsymbol{E}-\boldsymbol{P}\boldsymbol{\times}\boldsymbol{B}\right)d^{3}\boldsymbol{x}=\epsilon_{0}\int_{\partial V}\left[\boldsymbol{E}\boldsymbol{\otimes}\boldsymbol{E}+c^{2}\boldsymbol{B}\boldsymbol{\otimes}\boldsymbol{B}-\frac{1}{2}\boldsymbol{\mathbb{1}}_{3}\left(\boldsymbol{E}^{2}+c^{2}\boldsymbol{B}^{2}\right)\right]\boldsymbol{\cdot}d\boldsymbol{S}\label{classical-result}
	\end{multline}
	\end{widetext}
	where \(V\) is the volume of the dielectric, \(\partial V\) is the two dimensional region of free space on the boundary of \(V\), and \(d\boldsymbol{S}\) is an infinitesimal element of surface area directed outwards from the centre of the body.
	\par
	In the case where the body is instantaneously at rest, \(\dot{\boldsymbol{R}}=0\), then (\ref{classical-result}) can be reduced to the compact form,
	\begin{equation}
	M\ddot{\boldsymbol{R}}+\frac{\partial}{\partial t}\int_{V} \boldsymbol{\mathcal{P}}_{A} d^{3}\boldsymbol{x}=\int_{\partial V}\boldsymbol{\sigma}\boldsymbol{\cdot}d\boldsymbol{S}\label{force-1}
	\end{equation}
	where the momentum density is given by the Abraham expression~\cite{pfeifer2007},
	\begin{equation}
		\boldsymbol{\mathcal{P}}_{A}=\frac{1}{c^{2}}\boldsymbol{E}\boldsymbol{\times}\boldsymbol{H}\label{ab-mom}
	\end{equation}
	and the stress tensor is of the Maxwell form~\cite{volume2},
	\begin{equation}
		\boldsymbol{\sigma}=\epsilon_{0}\left[\boldsymbol{E}\boldsymbol{\otimes}\boldsymbol{E}+c^{2}\boldsymbol{B}\boldsymbol{\otimes}\boldsymbol{B}-\frac{1}{2}\boldsymbol{\mathbb{1}}_{3}(\boldsymbol{E}\boldsymbol{\cdot}\boldsymbol{E}+c^{2}\boldsymbol{B}\boldsymbol{\cdot}\boldsymbol{B})\right]\label{maxwell-stress}
	\end{equation}
	In deriving (\ref{force-1}) we have neglected a term proportional to the acceleration of the body times its spatial extent divided by \(c^{2}\).  This is required within the present context, for the acceleration must be small enough such that we do not have to worry about relativistic effects e.g.~\cite[Chapter 6]{wheeler1973}. 
	\par
	The expression for the acceleration of the centre of mass (\ref{force-1}) appears to agree with existing findings.  Consider the case of a short pulse entering an inhomogeneous dielectric.  During the time it propagates through the body, the field outside is approximately zero and expression (\ref{force-1}) suggests that the centre of mass should experience a force equal to minus the rate of change of the Abraham momentum within the medium.  This is in agreement with the result that the mechanical momentum of the pulse within the medium is associated with the Abraham expression~\cite{barnett2010,barnett2010b}.  Furthermore, if we consider an optical field interacting with the medium, and time average (\ref{force-1}) over an interval much longer than an optical cycle, then the force simply equals the integral of the averaged Maxwell stress tensor over the surface of the dielectric, which is the familiar expression used in studies of radiation pressure in free space~\cite{novotny2008}, and is equivalent to the average net macroscopic Lorentz force on the body (see~\cite{obukhov2003}). 
	\par
	Although (\ref{force-1}) is in itself not a surprising result, it is interesting that it can be derived without carrying out the procedure of constructing an energy momentum tensor as is done in e.g.~\cite{stallinga2006,philbin2011}.  Instead we started from an action that was constructed to give the correct constitutive relations for a moving medium and then varied this action with respect to the centre of mass velocity.  We might understand this procedure as being similar to the likes of~\cite{mansuripur2012a,mansuripur2012b}, where the Doppler shift in the constitutive relations of a medium is shown to be intimately connected with the force imparted by radiation.  This approach appears to be free from the ambiguity associated with the usual division of the full energy momentum tensor into `field' and `matter' contributions~\cite{stallinga2006}.
	\par
	If the reader performs the necessary manipulations to pass from (\ref{result-1}) to (\ref{force-1}) they may notice the divergence of the Minkowskii stress tensor, \(\boldsymbol{\sigma}_{\text{\tiny{M}}}=\boldsymbol{D}\boldsymbol{\otimes}\boldsymbol{E}+\boldsymbol{B}\boldsymbol{\otimes}\boldsymbol{H}-(1/2)\boldsymbol{\mathbb{1}}_{3}(\boldsymbol{D}\boldsymbol{\cdot}\boldsymbol{E}+\boldsymbol{B}\boldsymbol{\cdot}\boldsymbol{H})\) appears within the intermediate expressions (before divergences of terms containing the polarization and magnetization are dropped).  It is tempting to conclude that this tensor is somehow relevant to the force density within a medium.  However such a conclusion is not at all justified in the case of an isolated body in free space, given that once integrated over the volume of the medium there is no meaningful distinction between this intermediate tensor and (\ref{maxwell-stress}).  To investigate this point further, and to illustrate the application of our formalism to more than one dielectric body we now examine the electromagnetic force at the interface between two media.
%
%
	\subsection{Electromagnetic force at an interface between two dielectric media\label{interface-force}}
	\par
\begin{figure}[h]
	\includegraphics[width=5cm]{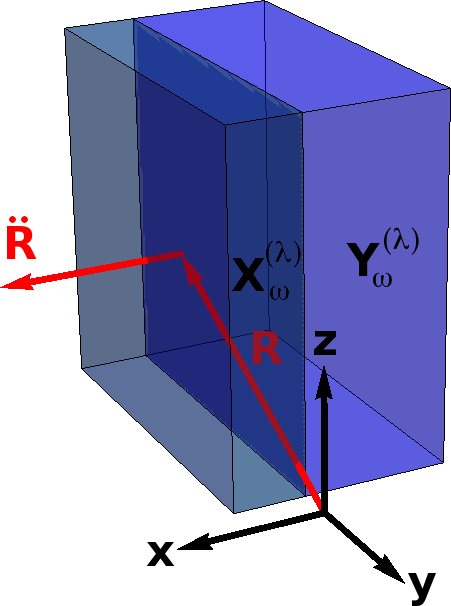}
	\caption{Two planar dielectric media are in contact over the plane \(x=0\) and the centre of mass of the first medium, \(\boldsymbol{R}\) (\(a>x>0\)) is free to move in response to the electromagnetic field.  The internal degrees of freedom of the medium occupying \(a>x>0\) are represented by the two sets of oscillators, \(\boldsymbol{X}_{\omega}^{(\lambda)}\), and those of the medium occupying \(x<0\) by \(\boldsymbol{Y}_{\omega}^{(\lambda)}\).\label{fig-1}}
	\end{figure}

	Consider the specific configuration shown in figure \ref{fig-1}, where we have two planar dielectric media instantaneously at rest and in contact over the \(x=0\) plane.  The first medium occupies the half space \(x<0\), and is taken to be of a mass so great that we can consider its position to be fixed.  The second medium occupies the space \(0<x<a\) and is allowed to move in response to its interaction with the electromagnetic field, which will include some contribution from the field at \(x=0\).
	\par
	To describe this situation in terms of an action we must introduce another pair of oscillator amplitudes into (\ref{modified-action}).  We denote these by \(\boldsymbol{Y}_{\omega}^{(\lambda)}\), and they describe the collective degrees of freedom of the medium occupying \(x<0\).  The action given in (\ref{modified-action}) is then equal to,
	\begin{multline}
		S^{\prime\prime}=S^{\prime}+\int d^{4}\boldsymbol{x}\int_{0}^{\infty}d\omega\bigg\{\sum_{\lambda}\frac{1}{2}\left(\dot{\boldsymbol{Y}}_{\omega}^{(\lambda)2}-\omega^{2}\boldsymbol{Y}_{\omega}^{(\lambda)2}\right)\\
		+\zeta(\omega)\boldsymbol{E}\boldsymbol{\cdot}\boldsymbol{Y}_{\omega}^{(\text{\tiny{E}})}+\xi(\omega)\boldsymbol{B}\boldsymbol{\cdot}\boldsymbol{Y}_{\omega}^{(\text{\tiny{B}})}\bigg\}-\int V(\boldsymbol{R}) d t \label{second_modified_action}
	\end{multline}
	Both media are assumed again isotropic and to be adequately represented by \(\epsilon\) and \(\mu\) in the rest frame.  The coupling constants, \(\zeta(\omega)\) and \(\xi(\omega)\) perform the same role as the \(\alpha(\omega)\) and \(\beta(\omega)\) in (\ref{polmag}), but are non--zero in the region \(x<0\) rather than \(0<x<a\).  The potential function \(V(\boldsymbol{R})\) represents the short range microscopic fields that prevent the two media from overlapping.  One obvious choice for \(V(\boldsymbol{R})\) would be an infinite potential barrier in the region \(\hat{\boldsymbol{x}}\boldsymbol{\cdot}\boldsymbol{R}<a/2\).
	\par
	As \(\boldsymbol{R}\) and \(\dot{\boldsymbol{R}}\) do not appear within the new terms in (\ref{second_modified_action}), the equation of motion (\ref{result-1}) is unchanged except for the addition of a term equal to \(-\boldsymbol{\nabla}_{\boldsymbol{R}}V(\boldsymbol{R})\) on the right hand side.  However, we should be careful not to make the identifications \(\boldsymbol{P}\) and \(\boldsymbol{M}\) within (\ref{result-1}) as given by (\ref{polmag}).  The total polarization and magnetization that appear with Maxwell's equations are now given by,
	\begin{align}
		\boldsymbol{P}&=\int_{0}^{\infty}\left[\alpha(\omega)\boldsymbol{X}_{\omega}^{(\text{\tiny{E}})}+\zeta(\omega)\boldsymbol{Y}_{\omega}^{(\text{\tiny{E}})}\right]d\omega=\boldsymbol{P}_{1}+\boldsymbol{P}_{2}\nonumber\\
		\boldsymbol{M}&=\int_{0}^{\infty}\left[\beta(\omega)\boldsymbol{X}_{\omega}^{(\text{\tiny{B}})}+\xi(\omega)\boldsymbol{Y}_{\omega}^{(\text{\tiny{B}})}\right]d\omega=\boldsymbol{M}_{1}+\boldsymbol{M}_{2}\label{total-P-M}
	\end{align}
	 we should therefore amend (\ref{result-1}) with \(\boldsymbol{P}\to\boldsymbol{P}_{1}\) and \(\boldsymbol{M}\to\boldsymbol{M}_{1}\).  After applying Maxwell's equations in the same manner as in the previous section we are left with
	 \begin{multline}
	 	M\ddot{\boldsymbol{R}}+\frac{\partial}{\partial t}\frac{1}{c^{2}}\int\boldsymbol{E}\boldsymbol{\times}\left(\boldsymbol{B}/\mu_{0}-\boldsymbol{M}_{1}\right)d^{3}\boldsymbol{x}+\boldsymbol{\nabla}_{\boldsymbol{R}}V(\boldsymbol{R})\\
		=\int\left[\boldsymbol{\nabla}\boldsymbol{\cdot}\boldsymbol{\sigma}+(\boldsymbol{\nabla}\boldsymbol{\cdot}\boldsymbol{P}_{2})\boldsymbol{E}-\left(\dot{\boldsymbol{P}}_{2}+\boldsymbol{\nabla}\boldsymbol{\times}\boldsymbol{M}_{2}\right)\boldsymbol{\times}\boldsymbol{B}\right]\label{force-2}
	 \end{multline}
	 To proceed we note that the following identity can be derived from Maxwell's equations,
	 \begin{multline}
	 \boldsymbol{\nabla}\boldsymbol{\cdot}\boldsymbol{\sigma}-\epsilon_{0}\frac{\partial}{\partial t}\left(\boldsymbol{E}\boldsymbol{\times}\boldsymbol{B}\right)=-(\boldsymbol{\nabla}\boldsymbol{\cdot}\boldsymbol{P})\boldsymbol{E}\\
	 +\left(\dot{\boldsymbol{P}}+\boldsymbol{\nabla}\boldsymbol{\times}\boldsymbol{M}\right)\boldsymbol{\times}\boldsymbol{B}\label{stress-identity}
	 \end{multline}
	 After applying (\ref{stress-identity}) to (\ref{force-2}) we find that the force on the centre of mass of the dielectric again obeys the equation of motion given by (\ref{force-1}), plus the short--ranged force due to the potential, \(V(\boldsymbol{R})\).  In this case too the theory predicts the centre of mass of the body to accelerate according to the net macroscopic Lorentz force, with the Abraham momentum relevant for the mechanical momentum exchanged with the electromagnetic field.  To prevent misunderstanding we should emphasise that this method of including dispersion and dissipation would not be valid for a body surrounded by a fluid - nor for the deformation of an elastic body - as the relevant action should include the local degrees of freedom of the body (e.g. the strain, or the fluid velocity), which we do not.  Therefore this finding is not immediately at odds with~\cite{pitaevskii2006,brevik2009}.  Nevertheless, in the remaining work we confine ourselves to the uncontroversial regime where the media are not in contact, and are surrounded by vacuum.
%
%
	\section{Quantum theory of motion}
	\par
	Having established that the action (\ref{modified-action}) reproduces the known results of the classical theory of radiation pressure, and mechanical momentum of light within media, we construct the corresponding quantum theory.  For the sake of simplicity we construct the Hamiltonian for a single dielectric body isolated in vacuum.  The case of two or more interacting bodies is not fundamentally different, and one can proceed as in the previous section.
	\par
	As usual, first find the canonical variables.  The canonical momentum for the electromagnetic field is given by,
	\begin{equation}
		\boldsymbol{\Pi}_{\boldsymbol{A}}=\frac{\partial\mathcal{L}}{\partial\dot{\boldsymbol{A}}}=-\epsilon_{0}\boldsymbol{E}-\boldsymbol{P}-\frac{1}{c^{2}}\dot{\boldsymbol{R}}\boldsymbol{\times}\boldsymbol{M}
	\end{equation}
	and for the oscillator fields,
	\begin{equation}
		\boldsymbol{\Pi}_{\boldsymbol{X}_{\omega}}^{(\lambda)}=\frac{\partial\mathcal{L}}{\partial\dot{\boldsymbol{X}}_{\omega}^{(\lambda)}}=\left(\frac{\partial}{\partial t}+\dot{\boldsymbol{R}}\boldsymbol{\cdot}\boldsymbol{\nabla}\right)\boldsymbol{X}_{\omega}^{(\lambda)}
	\end{equation}
	The canonical momentum associated with the centre of mass coordinate is given in (\ref{can-mom}).  Having obtained these variables, the Hamiltonian is then constricted in the usual way,
	\begin{multline}
		H=\int d^{3}\boldsymbol{x}\bigg(\boldsymbol{\Pi}_{\boldsymbol{A}}\boldsymbol{\cdot}\dot{\boldsymbol{A}}+\sum_{\lambda}\int_{0}^{\infty}d\omega \boldsymbol{\Pi}_{\boldsymbol{X}_{\omega}}^{(\lambda)}\boldsymbol{\cdot}\dot{\boldsymbol{X}}_{\omega}^{(\lambda)}\bigg)\\
		+\boldsymbol{p}\boldsymbol{\cdot}\dot{\boldsymbol{R}}-L^{\prime}\label{initial-hamiltonian}
	\end{multline}
	However, the relationship between the canonical variables and the mechanical ones is a little involved.  To cast the Hamiltonian entirely in terms of canonical variables we use the relation,
	\begin{equation}
		\left[\boldsymbol{\mathbb{1}}_{3}-\frac{\mu_{0}}{Mc^{2}}\int d^{3}\boldsymbol{x}\left(\boldsymbol{M}^{2}\boldsymbol{\mathbb{1}}_{3}-\boldsymbol{M}\boldsymbol{\otimes}\boldsymbol{M}\right)\right]\boldsymbol{\cdot}\dot{\boldsymbol{R}}=\frac{1}{M}\left(\boldsymbol{p}+\boldsymbol{\Gamma}\right)\label{velocity-momentum}
	\end{equation}
	where,
	\begin{multline}
		\boldsymbol{\Gamma}=\int d^{3}\boldsymbol{x}\big[\mu_{0}\boldsymbol{M}\boldsymbol{\times}\left(\boldsymbol{\Pi}_{\boldsymbol{A}}+\boldsymbol{P}\right)+\boldsymbol{P}\boldsymbol{\times}\left(\boldsymbol{\nabla}\boldsymbol{\times}\boldsymbol{A}\right)\\
		-\sum_{\lambda}\int_{0}^{\infty}d\omega\left(\boldsymbol{\nabla}\boldsymbol{\otimes}\boldsymbol{X}_{\omega}^{(\lambda)}\right)\boldsymbol{\cdot}\boldsymbol{\Pi}_{\boldsymbol{X}_{\omega}}^{(\lambda)}\big]\label{minimal-coupling}
	\end{multline}
	To simplify (\ref{velocity-momentum}), we might assume that the energy equal to the integral of the square of the magnetization over the medium times \(\mu_{0}\) is vanishingly small in comparison to the rest energy, \(Mc^{2}\).  From the perspective of classical physics this is permissible, but quantum mechanically the magnetization energy will contain a ground state contribution, which may in total not be small.  To account for this possibility, we assume an isotropic body and absorb this approximately constant energy---whatever it is---into the rest mass, \(M\) of the body.   Then (\ref{velocity-momentum}) becomes, \(\dot{\boldsymbol{R}}=(\boldsymbol{p}+\boldsymbol{\Gamma})/M\), and (\ref{initial-hamiltonian}) is given in terms of canonical variables by,
	\begin{multline}
	H=\frac{(\boldsymbol{p}+\boldsymbol{\Gamma})^{2}}{2M}+\int d^{3}\boldsymbol{x}\bigg[\frac{1}{2\epsilon_{0}}\left(\boldsymbol{\Pi}_{\boldsymbol{A}}+\boldsymbol{P}\right)^{2}+\frac{1}{2\mu_{0}}\boldsymbol{B}^{2}\\
	-\boldsymbol{M}\boldsymbol{\cdot}\boldsymbol{B}+\frac{1}{2}\sum_{\lambda}\int_{0}^{\infty}d\omega\left({\boldsymbol{\Pi}_{\boldsymbol{X}_{\omega}}^{(\lambda)}}^{2}+\omega^{2}{\boldsymbol{X}_{\omega}^{(\lambda)}}^{2}\right)\bigg]\label{classical-hamiltonian}
	\end{multline}
	where we have applied the constraint that \(\boldsymbol{\nabla}\boldsymbol{\cdot}\boldsymbol{\Pi}_{\boldsymbol{A}}=0\), and again dropped terms of order \((\dot{\boldsymbol{R}}/c)^{2}\).  Comparing (\ref{classical-hamiltonian}) to the Hamiltonian of~\cite{philbin2010} it is evident that the integral to the right of the kinetic energy term is the expression for the Hamiltonian of the electromagnetic field interacting with a dielectric at rest.  The centre of mass motion is coupled to the field and the dielectric by the term \(\boldsymbol{\Gamma}\), given in (\ref{minimal-coupling}).  This term is analogous to the vector potential in the Hamiltonian of a charged point particle interacting with the electromagnetic field.  However this `vector potential' is given by a quite complicated expression involving an integral of both field and material degrees of freedom over the whole body.
	\par
	Comparing (\ref{classical-hamiltonian}) with the expressions of~\cite{law1995,cheung2012} we see that Law and Cheung's result---that a dielectric body couples `minimally' to the electromagnetic field---also holds when dispersion and dissipation are accounted for.  Given this form of the coupling, we expect Aharonov--Bohm type effects to be present within the behaviour of the centre of mass, with the semiclassical phase shift given by,
	\begin{equation}
		\Delta\phi=-\frac{1}{\hbar}\oint\boldsymbol{\Gamma}\boldsymbol{\cdot}d\boldsymbol{R}\label{ab-phase}
	\end{equation}
	Presumably this phase is very difficult to directly observe within an interference pattern---e.g. in the same manner as for neutrons~\cite{sangster1995}---and some thought must go into the indirect implications.  We note that the analogue of this effect was also found in~\cite{cheung2012} for the case where dissipation and dispersion are neglected.  The form of the effective vector potential given in (\ref{minimal-coupling}) contains terms that may be identified as the equivalent of the Aharonov--Casher~\cite{aharonov1984} and He--McKellar--Wilkens~\cite{he1993,wilkens1994} contributions for an extended object.  However, we find an additional term is also present due to our continuum description of dispersion and dissipation.  This does not arise in the case where the object is treated as a discrete collection of oscillators~\cite{horsley2008}, but seems to be necessary in macroscopic electromagnetism.
	\par
	The Hamiltonian given in (\ref{classical-hamiltonian}) is bounded from below.  In~\cite{horsley2012} it was found that if we attempt to construct a quantum field theory for electromagnetism interacting with a uniformly moving dielectric then the Hamiltonian is extremely odd and, without including the effects of spatial dispersion, unbounded from below.  We again interpret these odd features as arising from a failure to include the dynamics of the centre of mass motion, which are inseparable from those of the field.  The Hamiltonian of~\cite{horsley2012} should therefore be understood as an approximate description of the physical situation, with (\ref{classical-hamiltonian}) being more fundamental.
	\par
	The quantum mechanical Hamiltonian is obtained from (\ref{classical-hamiltonian}) in the usual way.  The canonical variables become operators satisfying the canonical commutation relations.  The centre of mass operators obey, 
	\begin{equation}
		\left[\hat{\boldsymbol{R}},\hat{\boldsymbol{p}}\right]=i\hbar\boldsymbol{\mathbb{1}}_{3}\label{pos-com}
	\end{equation}
	and the electromagnetic field operators satisfy,
	\begin{equation}
		\left[\hat{\boldsymbol{A}}(\boldsymbol{x},t),\hat{\boldsymbol{\Pi}}_{\boldsymbol{A}}(\boldsymbol{x}^{\prime},t)\right]=i\hbar\boldsymbol{\delta}_{\perp}(\boldsymbol{x}-\boldsymbol{x}^{\prime})\label{em-com}
	\end{equation}
	where \(\boldsymbol{\delta}_{\perp}(\boldsymbol{x}-\boldsymbol{x}^{\prime})\) is the transverse delta function~\cite{loudon1983}.  Finally, the commutation relations for the oscillator field operators are,
	\begin{multline}
		\left[\hat{\boldsymbol{X}}_{\omega}^{(\lambda)}(\boldsymbol{x},t),\hat{\boldsymbol{\Pi}}_{\boldsymbol{X}_{\omega^{\prime}}}^{(\lambda^{\prime})}(\boldsymbol{x}^{\prime},t)\right]\\
		=i\hbar\boldsymbol{\mathbb{1}}_{3}\delta_{\lambda\lambda^{\prime}}\delta(\omega-\omega^{\prime})\delta^{(3)}(\boldsymbol{x}-\boldsymbol{x}^{\prime})\label{osc-com}
	\end{multline}
	The Hamiltonian operator then takes a very similar form to (\ref{classical-hamiltonian}),
	\begin{equation}
		\hat{H}=\frac{\left(\hat{\boldsymbol{p}}+\hat{\boldsymbol{\Gamma}}\right)^{2}}{2M}+\hat{H}_{\text{\tiny{P}}}(\hat{\boldsymbol{R}})\label{quantum-hamiltonian}
	\end{equation}
	where,
	\begin{widetext}
	\begin{equation}
		\hat{H}_{\text{\tiny{P}}}(\hat{\boldsymbol{R}})=\int d^{3}\boldsymbol{x}\bigg[\frac{1}{2\epsilon_{0}}\left(\hat{\boldsymbol{\Pi}}_{\boldsymbol{A}}+\hat{\boldsymbol{P}}\right)^{2}+\frac{1}{2\mu_{0}}\left(\boldsymbol{\nabla}\boldsymbol{\times}\hat{\boldsymbol{A}}\right)^{2}
		-\hat{\boldsymbol{M}}\boldsymbol{\cdot}\left(\boldsymbol{\nabla}\boldsymbol{\times}\hat{\boldsymbol{A}}\right)+\frac{1}{2}\sum_{\lambda}\int_{0}^{\infty}d\omega\left(\hat{\boldsymbol{\Pi}}_{\boldsymbol{X}_{\omega}}^{(\lambda)2}+\omega^{2}\hat{\boldsymbol{X}}_{\omega}^{(\lambda)2}\right)\bigg]\label{Hp}
	\end{equation}
	\end{widetext}
	the only distinction being that a symmetric ordering of the oscillator field operators and their associated canonical momenta must be chosen in the \(\hat{\boldsymbol{\Gamma}}\) operator,
	\begin{widetext}
	\begin{equation}
		\hat{\boldsymbol{\Gamma}}=\int d^{3}\boldsymbol{x}\bigg\{\mu_{0}\hat{\boldsymbol{M}}\boldsymbol{\times}\left(\hat{\boldsymbol{\Pi}}_{\boldsymbol{A}}+\hat{\boldsymbol{P}}\right)+\hat{\boldsymbol{P}}\boldsymbol{\times}(\boldsymbol{\nabla}\boldsymbol{\times}\hat{\boldsymbol{A}})
		-\frac{1}{2}\sum_{\lambda}\int_{0}^{\infty}d\omega\left[(\boldsymbol{\nabla}\boldsymbol{\otimes}\hat{\boldsymbol{X}}_{\omega}^{(\lambda)})\boldsymbol{\cdot}\hat{\boldsymbol{\Pi}}_{\boldsymbol{X}_{\omega}}^{(\lambda)}+\hat{\boldsymbol{\Pi}}_{\boldsymbol{X}_{\omega}}^{(\lambda)}\boldsymbol{\cdot}(\hat{\boldsymbol{X}}_{\omega}^{(\lambda)}\boldsymbol{\otimes}\overleftarrow{\boldsymbol{\nabla}})\right]\bigg\}
	\end{equation}
	\end{widetext}
	The part of the (\ref{quantum-hamiltonian}) given by (\ref{Hp}) is identical to that of a stationary dielectric at a position given by \(\hat{\boldsymbol{R}}\), and may be `diagonalized' into `normal modes' (`polaritons' in the sense of~\cite{huttner1992}) of the combined system of the field and the internal degrees of freedom of the dielectric~\cite{philbin2010}.  The total `polariton' energy can then be written in terms of a number operator.  The structure of (\ref{quantum-hamiltonian}) is made especially interesting by this fact that the number operator depends on the centre of mass operator, \(\hat{\boldsymbol{R}}\): for each possible position of the dielectric body we have a different basis of `polariton' number states.\\
	\par
	From this discussion, it is not clear what the ground state of (\ref{quantum-hamiltonian}) is.  We might expect that the ground state would correspond to a completely de--localized centre of mass coordinate and zero `polaritons', i.e. \(|\psi_{0}\rangle=N_{0}|0\rangle_{\boldsymbol{R}}\), where \(N_{0}\) is a normalization factor, and \(|0\rangle_{\boldsymbol{R}}\) is the state of zero `polaritons' when counted using the basis associated with the centre of mass being at \(\boldsymbol{R}\).  However, this expression for \(|\psi_{0}\rangle\) does not appear to be an eigenstate of (\ref{quantum-hamiltonian}), due to the fact that \(\hat{\boldsymbol{p}}|0\rangle_{\boldsymbol{R}}\neq0\).  We leave the problem of finding these eigenstates for future work, and to conclude we apply (\ref{quantum-hamiltonian}) to the spreading of a localised centre of mass when we initially have zero `polaritons'.\\
%
%
	\subsection{Quantum motion of an initially localised centre of mass\label{wave-packet}}
	\par
\begin{figure}
		\includegraphics[width=8cm]{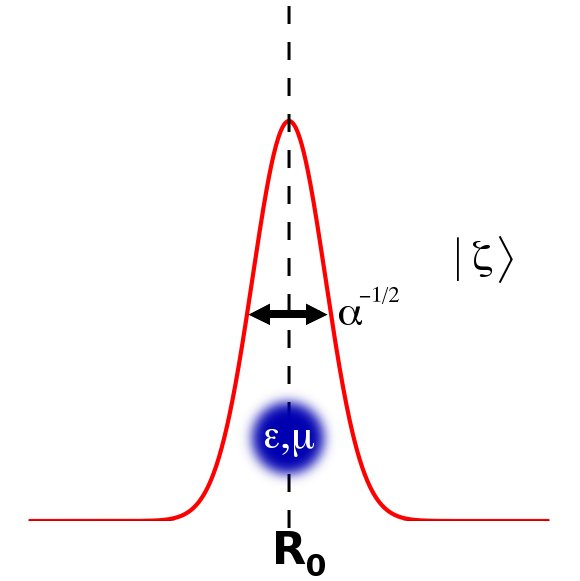}
		\caption{A small object, characterized by \(\epsilon\) and \(\mu\) satisfying the Kramers--Kr\"onig relations is prepared so that the probability density for the centre of mass is spread over a width, \(\alpha^{-1/2}\) (c.f. (\ref{initial-state})).  The coupled system of field and dielectric is initially in the state \(|\zeta\rangle\), and we investigate the effect of the coupling between this system and the centre of mass on the time evolution of the wave-packet.\label{figure-2}}
	\end{figure}

	In order to glean some understanding of the quantum centre of mass motion, we consider a state where at a time \(t=0\) the centre of mass has been prepared so that it is sharply localised around some mean position, \(\boldsymbol{R}_{0}\) (see figure~\ref{figure-2}).  If the mass of the object is large enough then the characteristic spreading time of a wavepacket of width \(\alpha^{-1/2}\), \(T_{s}=M/\hbar\alpha\) can be long compared to the dynamics of the field.  For example, were we to know the initial centre of mass position to within \(10^{-9}\,\text{m}\) (\(\alpha\sim10^{18}\,\text{m}^{2}\)), then a system of \(\sim10^{12}\) atoms (\(M\sim10^{-14}\,\text{kg}\)) would exhibit \(T_{s}\sim10^{2}\,\text{s}\).  Therefore, in the regime of macroscopic electromagnetism, a good approximation is to expand (\ref{quantum-hamiltonian}) around \(\boldsymbol{R}_{0}\)~\cite{law1995}, a significant time being required before terms of order \((\boldsymbol{R}-\boldsymbol{R}_{0})^{2}\) and higher become relevant to the time evolution of \(|\psi\rangle\)~\footnote{One might criticise the expansion (\ref{H_approx}) on the basis that the high frequency modes of the field vary in a non--linear manner over the extent of the wave--packet. We can avoid this complication if we consider the dispersion of the medium to be such that it is transparent at such high frequencies.}.  Such a first order expansion of (\ref{quantum-hamiltonian}) yields,
	\begin{widetext}
	\begin{multline}
		\hat{H}\sim\frac{\left(\hat{\boldsymbol{p}}+\hat{\boldsymbol{\Gamma}}_{0}+\dots\right)^{2}}{2M}+(\boldsymbol{R}-\boldsymbol{R}_{0})\boldsymbol{\cdot}\int d^{3}\boldsymbol{x}\bigg[\frac{1}{\epsilon_{0}}(\boldsymbol{\nabla}_{\boldsymbol{R}}\boldsymbol{\otimes}\hat{\boldsymbol{P}})_{0}\boldsymbol{\cdot}(\hat{\boldsymbol{\Pi}}_{\boldsymbol{A}}+\hat{\boldsymbol{P}}_{0})-(\boldsymbol{\nabla}_{\boldsymbol{R}}\boldsymbol{\otimes}\hat{\boldsymbol{M}})_{0}\boldsymbol{\cdot}\left(\boldsymbol{\nabla}\boldsymbol{\times}\hat{\boldsymbol{A}}\right)\bigg]+\hat{H}_{0}\label{H_approx}
	\end{multline}
	\end{widetext}
	where the zero subscript indicates that the quantity is evaluated at \(\boldsymbol{R}=\boldsymbol{R}_{0}\), \(\hat{H}_{0}=\hat{H}_{\text{\tiny{P}}}(\boldsymbol{R}_{0})\), and the ellipsis denote further terms in the expansion of \(\hat{\boldsymbol{\Gamma}}\) around \(\boldsymbol{R}_{0}\).  As mentioned in the previous section, we can diagonalise \(\hat{H}_{0}\) into `polariton' modes with the transformation of~\cite{philbin2010} (see appendix~\ref{appendix-A}), which gives
	\begin{multline}
		\hat{H}_{0}=\frac{1}{2}\sum_{\lambda}\int d^{3}\boldsymbol{x}\int_{0}^{\infty}d\omega \hbar\omega\hat{\boldsymbol{C}}_{\omega}^{(\lambda)\dagger}(\boldsymbol{x};\boldsymbol{R}_{0})\boldsymbol{\cdot}\hat{\boldsymbol{C}}_{\omega}^{(\lambda)}(\boldsymbol{x};\boldsymbol{R}_{0})\\
		+\text{h.c.}\label{H0}
	\end{multline}
	We shall often assume that \(\hat{H}_{0}\) takes the above diagonalised form, and that the remaining `polariton' (non--mechanical) degrees of freedom in (\ref{H_approx}) are expanded in terms of the \(\hat{\boldsymbol{C}}_{\omega}\) operators.
	\par
	Before proceeding with the calculation, we choose to transform (\ref{H_approx}) into a different representation.  Due to the independence of (\ref{H0}) from \(\boldsymbol{R}\) we can quite simply cast (\ref{H_approx}) into an interaction representation via the unitary transformation,
	\[
		\hat{U}_{1}(t)=\exp{\left(-\frac{it}{\hbar}\hat{H}_{0}\right)}
	\]
	where we identify,
	\begin{align}
	|\psi\rangle&=\hat{U}_{1}|\varphi\rangle\nonumber\\
	\hat{H}&=\hat{U}_{1}\left(\hat{H}_{I}+\hat{H}_{0}\right)\hat{U}_{1}^{\dagger}
	\end{align}
	We then find the usual result, \(\hat{H}_{I}|\varphi\rangle=i\hbar\partial|\varphi\rangle/\partial t\), where \(\hat{H}_{I}=\hat{U}_{1}^{\dagger}(\hat{H}-\hat{H}_{0})\hat{U}_{1}\), and the effect of the unitary transformation on (\ref{H_approx}) is to make the electromagnetic field, polarization, and magnetization operators all time dependent, via the substitution \(\hat{\boldsymbol{C}}_{\omega}\to\hat{\boldsymbol{C}}_{\omega}\exp{(-i\omega t)}\).  In order to simplify the situation further, we make a second unitary transformation,
	\begin{equation}
		\hat{U}_{2}(\boldsymbol{R},t)=\exp\bigg[-\frac{i}{\hbar}(\boldsymbol{R}-\boldsymbol{R}_{0})\boldsymbol{\cdot}\hat{\boldsymbol{\Gamma}}_{0}\bigg]\label{U2}
	\end{equation}
	where we write, \(|\varphi\rangle=\hat{U}_{2}|\chi\rangle\), and \(\hat{H}_{I}=\hat{U}_{2}\hat{{H}}^{\prime}_{I}\hat{U}_{2}^{\dagger}+i\hbar(\partial\hat{U}_{2}/\partial t)\hat{U}_{2}^{\dagger}\).  This is the representation of the Schr\"odinger equation which we apply to the motion of the wave--packet: \(\hat{H}^{\prime}_{I}|\chi\rangle=i\hbar\partial|\chi\rangle/\partial t\), the approximate Hamiltonian being given by,
	\begin{widetext}
	\begin{multline}
		\hat{H}^{\prime}_{I}(t)=\frac{\hat{\boldsymbol{p}}^{2}}{2M}+(\boldsymbol{R}-\boldsymbol{R}_{0})\boldsymbol{\cdot}\left\{\int d^{3}\boldsymbol{x}\left[\frac{1}{\epsilon_{0}}(\boldsymbol{\nabla}_{\boldsymbol{R}}\boldsymbol{\otimes}\hat{\boldsymbol{P}})_{0}\boldsymbol{\cdot}(\hat{\boldsymbol{\Pi}}_{\boldsymbol{A}}+\hat{\boldsymbol{P}}_{0})-(\boldsymbol{\nabla}_{\boldsymbol{R}}\boldsymbol{\otimes}\hat{\boldsymbol{M}})_{0}\boldsymbol{\cdot}\left(\boldsymbol{\nabla}\boldsymbol{\times}\hat{\boldsymbol{A}}\right)\right]-\frac{\partial\hat{\boldsymbol{\Gamma}}_{0}}{\partial t}\right\}\\
		+\frac{1}{2M}\left[\hat{\boldsymbol{p}}\boldsymbol{\cdot}(\hat{\boldsymbol{\Gamma}}\boldsymbol{\otimes}\overleftarrow{\boldsymbol{\nabla}}_{\boldsymbol{R}})_{0}\boldsymbol{\cdot}(\boldsymbol{R}-\boldsymbol{R}_{0})+(\boldsymbol{R}-\boldsymbol{R}_{0})\boldsymbol{\cdot}(\boldsymbol{\nabla}_{\boldsymbol{R}}\boldsymbol{\otimes}\hat{\boldsymbol{\Gamma}})_{0}\boldsymbol{\cdot}\hat{\boldsymbol{p}}\right]\label{Hintnew}
	\end{multline}
	\end{widetext}
	where all terms have been truncated to first order in \(\boldsymbol{R}-\boldsymbol{R}_{0}\), and the Hermitian part of the resulting expression has been taken.  When evaluating the average acceleration of the wave--packet, the second line of (\ref{Hintnew}) will in general give something distinct from the corresponding classical result.  This is because the ordering of the position and momentum operators matters.  However we do not investigate such effects here.
	\par
	Our attention now turns more precisely the dynamics of a wave--packet initially prepared in the state illustrated in figure~\ref{figure-2},
	\begin{equation}
		|\chi(t=0)\rangle=\left(\frac{\alpha}{\pi}\right)^{3/4}e^{-\frac{\alpha}{2}(\boldsymbol{R}-\boldsymbol{R}_{0})^{2}}|\zeta\rangle\label{initial-state}
	\end{equation}
	where \(|\zeta\rangle\) is an eigenstate of the \(\hat{\boldsymbol{C}}^{(\lambda)}_{\omega}(\boldsymbol{x},\boldsymbol{R}_{0})\) operators.  It is reasonably straightforward to show that for such a state, the average position is subject to a mean instantaneous acceleration that can be interpreted as the analogue of the classical result (\ref{force-1}).  Firstly, via the relation \(\langle\dot{\boldsymbol{R}}\rangle=-(i/\hbar)\langle[\boldsymbol{R},\hat{H}^{\prime}_{I}]\rangle\) we find the average velocity of the state (\ref{initial-state}) to be zero,
	\begin{equation}
		\langle\dot{\boldsymbol{R}}\rangle=\frac{1}{M}\left\langle\chi\right|\left[\hat{\boldsymbol{p}}+(\boldsymbol{R}-\boldsymbol{R}_{0})\boldsymbol{\cdot}(\boldsymbol{\nabla}_{\boldsymbol{R}}\boldsymbol{\otimes}\hat{\boldsymbol{\Gamma}})_{0}\right]\left|\chi\right\rangle=0\label{noV}
	\end{equation}
	Introducing the quantity, \(\hat{\boldsymbol{V}}=M^{-1}[\hat{\boldsymbol{p}}+(\boldsymbol{R}-\boldsymbol{R}_{0})\boldsymbol{\cdot}(\boldsymbol{\nabla}_{\boldsymbol{R}}\boldsymbol{\otimes}\hat{\boldsymbol{\Gamma}})_{0}]\), the average mechanical force can be similarly computed from the relation, \(M\langle\ddot{\boldsymbol{R}}\rangle=-(iM/\hbar)\langle[\hat{\boldsymbol{V}},\hat{H}_{I}^{\prime}]\rangle\), which yields,
	\begin{widetext}
	\begin{align}
		M\langle\ddot{\boldsymbol{R}}\rangle&=\left\langle\chi\right|\left\{\int d^{3}\boldsymbol{x}\left[\boldsymbol{\nabla}_{\boldsymbol{R}}\boldsymbol{\otimes}\left(\hat{\boldsymbol{P}}+\frac{1}{c^{2}}\hat{\boldsymbol{V}_{0}}\boldsymbol{\times}\hat{\boldsymbol{M}}\right)\boldsymbol{\cdot}\hat{\boldsymbol{E}}^{\prime}+\boldsymbol{\nabla}_{\boldsymbol{R}}\boldsymbol{\otimes}\left(\hat{\boldsymbol{M}}-\hat{\boldsymbol{V}}_{0}\boldsymbol{\times}\hat{\boldsymbol{P}}\right)\boldsymbol{\cdot}\hat{\boldsymbol{B}}\right]_{0}+\frac{\partial\hat{\boldsymbol{\Gamma}}_{0}}{\partial t}+\hat{\boldsymbol{V}}\boldsymbol{\cdot}(\boldsymbol{\nabla}_{\boldsymbol{R}}\boldsymbol{\otimes}\hat{\boldsymbol{\Gamma}})_{0}\right\}\left|\chi\right\rangle\nonumber\\
		&=\left\langle\chi\right|\left\{\int d^{3}\boldsymbol{x}\left[\left(\boldsymbol{\nabla}_{\boldsymbol{R}}\boldsymbol{\otimes}\hat{\boldsymbol{P}}\right)\boldsymbol{\cdot}\hat{\boldsymbol{E}}+\left(\boldsymbol{\nabla}_{\boldsymbol{R}}\boldsymbol{\otimes}\hat{\boldsymbol{M}}\right)\boldsymbol{\cdot}\hat{\boldsymbol{B}}\right]_{0}+\frac{\partial\hat{\boldsymbol{\Gamma}}_{0}}{\partial t}\right\}\left|\chi\right\rangle\label{quantum-force}
	\end{align}
	\end{widetext}
	where we have defined the quantities, \(\hat{\boldsymbol{E}}^{\prime}=-\epsilon_{0}^{-1}[\hat{\boldsymbol{\Pi}}_{\boldsymbol{A}}+\hat{\boldsymbol{P}}_{0}+c^{-2}\hat{\boldsymbol{V}}\boldsymbol{\times}\hat{\boldsymbol{M}}_{0}]\), \(\hat{\boldsymbol{E}}=-\epsilon_{0}^{-1}[\hat{\boldsymbol{\Pi}}_{\boldsymbol{A}}+\hat{\boldsymbol{P}}_{0}]\), and \(\hat{\boldsymbol{B}}=\boldsymbol{\nabla}\boldsymbol{\times}\hat{\boldsymbol{A}}\), and applied the result (\ref{noV}).  The velocity dependent contribution to the electric field operator drops out in the second line of (\ref{quantum-force}), due to the form of (\ref{initial-state}).  Note that the expression on the first line of (\ref{quantum-force}) is the operator equivalent of (\ref{lagrange-eqn}).  Besides assuming that the field does not vary significantly over the extent of the wave--packet, the derivation of (\ref{quantum-force}) only relies on the fact that (\ref{initial-state}) is both separable and rotationally symmetric around the point \(\boldsymbol{R}_{0}\).  Applying the equations of motion of the operators---\(\partial\hat{\boldsymbol{\Gamma}}_{0}/\partial t=-(i/\hbar)[\hat{\boldsymbol{\Gamma}}_{0},\hat{H}_{0}]\)---to the time derivative in (\ref{quantum-force}), and performing several integrations by parts, one obtains the result,
	\begin{equation}
		M\langle\ddot{\boldsymbol{R}}\rangle=\left\langle\chi\right|\left(\int_{\partial V} \hat{\boldsymbol{\sigma}}\boldsymbol{\cdot}d\boldsymbol{S}-\frac{1}{c^{2}}\frac{\partial}{\partial t}\int_{V}\hat{\boldsymbol{\mathcal{P}}}_{A}d^{3}\boldsymbol{x}\right)|\chi\rangle\label{quantum-wave-packet-force}
	\end{equation}
	where \(\hat{\boldsymbol{\mathcal{P}}}_{A}=\hat{\boldsymbol{E}}\boldsymbol{\times}\hat{\boldsymbol{H}}\), \(\hat{\boldsymbol{H}}=\hat{\boldsymbol{B}}/\mu_{0}-\hat{\boldsymbol{M}}_{0}\), and \(\hat{\boldsymbol{\sigma}}=\epsilon_{0}[\hat{\boldsymbol{E}}\boldsymbol{\otimes}\hat{\boldsymbol{E}}+c^{2}\hat{\boldsymbol{B}}\boldsymbol{\otimes}\hat{\boldsymbol{B}}-(1/2)\boldsymbol{\mathbb{1}}_{3}(\hat{\boldsymbol{E}}^{2}+c^{2}\hat{\boldsymbol{B}}^{2})]\).  It is therefore clear that the average position of an initially stationary and well localised centre of mass is subject to the expectation value of the operator equivalent of (\ref{force-1})~\footnote{When the object is in motion, the situation is in general significantly more complicated, not least because the velocity operator \(\hat{\boldsymbol{V}}\) does not commute with the field operators.}. This is comforting from the point of verifying that this theory gives sensible answers, but is wholly anticipated given that expectation values are known to exhibit behaviour consistent with classical equations of motion.
	\par
	We now move from a discussion of average properties computed from the state, to the state itself.  To solve the Schr\"odinger equation we proceed pertubatively, the state being expanded as \(|\chi(t)\rangle=|\chi^{(0)}(t)\rangle+|\chi^{(1)}(t)\rangle\), obeying the zeroth and first order relations,
	\begin{align}
		\frac{\hat{\boldsymbol{p}}^{2}}{2M}|\chi^{(0)}(t)\rangle-i\hbar\frac{\partial|\chi^{(0)}(t)\rangle}{\partial t}&=0\nonumber\\
		\frac{\hat{\boldsymbol{p}}^{2}}{2M}|\chi^{(1)}(t)\rangle-i\hbar\frac{\partial|\chi^{(1)}(t)\rangle}{\partial t}&=-\Delta\hat{H}(t)|\chi^{(0)}(t)\rangle\label{perturbation-exp}
	\end{align}
	where \(\hat{H}_{I}^{\prime}(t)=\hat{\boldsymbol{p}}^{2}/2M + \Delta\hat{H}(t)\).  The idea is that, as alluded to at the beginning of this section, if the spreading time of the wave--packet is long then over some relatively extended time interval the object will remain well localised around \(\boldsymbol{R}_{0}\).  With some approximation we can then consider the first power of \(\boldsymbol{R}-\boldsymbol{R}_{0}\) to dominate the time evolution of \(|\chi\rangle\).  Applying the condition, \(|\chi(t=0)\rangle=|\chi^{(0)}(t=0)\rangle\), the first of (\ref{perturbation-exp}) has the solution,
	\begin{equation}
		|\chi^{(0)}(t)\rangle=\left(\frac{\sqrt{\alpha/\pi}}{1+i\hbar\alpha t/M}\right)^{3/2}e^{-\frac{\alpha(\boldsymbol{R}-\boldsymbol{R}_{0})^{2}}{2\left(1+i\hbar\alpha t/M\right)}}|\zeta\rangle\label{zeroth-order-result}
	\end{equation}
	After inserting (\ref{zeroth-order-result}) into the second line of (\ref{perturbation-exp}) we can use the free particle Green function, \(K(\boldsymbol{R},\boldsymbol{R}^{\prime};T)\) (\(T=t-t^{\prime}\)) to find an expression for \(|\chi^{(1)}(t)\rangle\).  The free particle Green function is a solution of,
	\[
		\left[\frac{\hat{\boldsymbol{p}}^{2}}{2M}-i\hbar\frac{\partial}{\partial t}\right]K(\boldsymbol{R},\boldsymbol{R}^{\prime};T)=-i\hbar\delta^{(3)}(\boldsymbol{R}-\boldsymbol{R}^{\prime})\delta(t-t^{\prime})
	\]
	and has the explicit form,
	\begin{equation}
		K(\boldsymbol{R},\boldsymbol{R}^{\prime};T)=\left(\frac{M}{2\pi i\hbar T}\right)^{3/2}e^{\frac{iM}{2\hbar T}(\boldsymbol{R}-\boldsymbol{R}^{\prime})^{2}}\Theta(T)\label{propagator}
	\end{equation}\\
	which is the retarded Green function.  In terms of (\ref{propagator}), the solution to the second line of (\ref{perturbation-exp}) is,
	\begin{multline}
		|\chi^{(1)}(t)\rangle=-\frac{i}{\hbar}\int_{0}^{t}d t^{\prime}\int d^{3}\boldsymbol{R}^{\prime}K(\boldsymbol{R},\boldsymbol{R}^{\prime};T)\\
		\times\Delta\hat{H}(\boldsymbol{R}^{\prime},t^{\prime})|\chi^{(0)}(\boldsymbol{R}^{\prime},t^{\prime})\rangle\label{formal-solution}
	\end{multline}
	The spatial integral in (\ref{formal-solution}) is Gaussian, which can be evaluated using the usual results.  Inserting (\ref{Hintnew}), (\ref{zeroth-order-result}) and (\ref{propagator}) into (\ref{formal-solution}), and performing this integral gives,
	\begin{widetext}
	\begin{equation}
		|\chi^{(1)}(t)\rangle=\frac{i}{\hbar}\int_{0}^{t}dt^{\prime}\left\{\frac{i\hbar}{2M}(\boldsymbol{\nabla_{\boldsymbol{R}}}\boldsymbol{\cdot}\hat{\boldsymbol{\Gamma}})_{0}-\left(\frac{1+i\hbar\alpha t^{\prime}/M}{1+i\hbar\alpha t/M}\right)(\boldsymbol{R}-\boldsymbol{R}_{0})\boldsymbol{\cdot}\left[\frac{1}{c^{2}}\int_{V} d^{3}\boldsymbol{x}\frac{\partial\hat{\boldsymbol{\mathcal{P}}}_{A}}{\partial t^{\prime}}-\int_{\partial V} d\boldsymbol{S}\boldsymbol{\cdot}\hat{\boldsymbol{\sigma}}\right]\right\}|\chi^{(0)}(t)\rangle\rangle\label{first-order-correction}
	\end{equation}
	\end{widetext}
	where we have applied the field operator equations of motion in the same way as in the derivation of (\ref{quantum-wave-packet-force}), and kept to first order in \(\boldsymbol{R}-\boldsymbol{R}_{0}\).  We note the appearance of \((\boldsymbol{\nabla}_{\boldsymbol{R}}\boldsymbol{\cdot}\hat{\boldsymbol{\Gamma}})_{0}\) within the curly brackets in (\ref{first-order-correction}).  Such a term is certainly peculiar to quantum mechanics, where it arises because position and momentum variables do not commute.  This term stems from the fact that the canonical momentum of the centre of mass has both a contribution from the velocity of the medium and the field degrees of freedom.  The second term in curly brackets is the operator equivalent of the classical radiation pressure force investigated in section~\ref{classical-radiation-pressure}.  Being multiplied by \(\boldsymbol{R}-\boldsymbol{R}_{0}\), this serves to shift the centre of the wave-packet in accordance with (\ref{quantum-wave-packet-force}).  However, besides shifting the average position of the wave--packet and leaving the field in the \(|\zeta\rangle\) state, this term also creates pairs of `polariton' excitations as the wave--packet spreads (i.e. there are pairs of `polariton' creation operators within (\ref{first-order-correction})), and this will cause a fluctuation of the form of the packet, as well as a change in the state of the field.  As a final calculation we look to make this discussion more concrete and discern the regime of parameters where the above effect is evident.
	\par
	The field operators in (\ref{first-order-correction}) are expanded in terms of the \(\hat{\boldsymbol{C}}_{\omega}\) and \(\boldsymbol{C}^{\dagger}_{\omega}\) operators given in appendix~\ref{appendix-A}.  We also specialise to the case where \(|\zeta\rangle=|0\rangle\), which is the state for which \(\hat{\boldsymbol{C}}_{\omega}(\boldsymbol{x})|0\rangle=0\), for any choice of annihilation operator.  Applying (\ref{C1}--\ref{C2}) we find that the operator equivalent of the stress tensor acting on the vacuum state gives a sum of two identifiable terms,
	\begin{widetext}
	\begin{multline}
		\hat{\boldsymbol{\sigma}}(\boldsymbol{x})|0\rangle=\epsilon_{0}\int_{0}^{\infty}d\Omega\int_{0}^{\infty}d\Omega^{\prime}e^{i(\Omega+\Omega^{\prime})t^{\prime}}\left\{\hat{\boldsymbol{\upsilon}}(\boldsymbol{x},\Omega,\Omega^{\prime})-\frac{1}{2}\boldsymbol{\mathbb{1}}_{3}\text{Tr}\left[\hat{\boldsymbol{\upsilon}}(\boldsymbol{x},\Omega,\Omega^{\prime})\right]\right\}|0\rangle\\
		+\frac{\hbar\epsilon_{0}}{\pi}\lim_{\boldsymbol{x}\to\boldsymbol{x}^{\prime}}\int_{0}^{\infty}d\Omega\,\text{Im}\left\{\boldsymbol{\tau}(\boldsymbol{x},\boldsymbol{x}^{\prime}\Omega)-\frac{1}{2}\boldsymbol{\mathbb{1}}_{3}\text{Tr}\left[\boldsymbol{\tau}(\boldsymbol{x},\boldsymbol{x}^{\prime},\Omega)\right]\right\}|0\rangle\label{stress-vacuum}
	\end{multline}
	\end{widetext}
	where in (\ref{stress-vacuum}) we have defined
	\begin{equation}
		\boldsymbol{\tau}(\boldsymbol{x},\boldsymbol{x}^{\prime},\Omega)=\Omega^{2}\boldsymbol{G}(\boldsymbol{x},\boldsymbol{x}^{\prime};\Omega)
		+c^{2}\boldsymbol{\nabla}\boldsymbol{\times}\boldsymbol{G}(\boldsymbol{x},\boldsymbol{x}^{\prime};\Omega)\boldsymbol{\times}\overleftarrow{\boldsymbol{\nabla}}^{\prime}\label{tau}
	\end{equation}
	and
	\begin{multline}
		\hat{\boldsymbol{\upsilon}}(\boldsymbol{x},\Omega,\Omega^{\prime})=\hat{\boldsymbol{\mathcal{E}}}^{\dagger}(\boldsymbol{x},\Omega)\boldsymbol{\otimes}\hat{\boldsymbol{\mathcal{E}}}^{\dagger}(\boldsymbol{x},\Omega^{\prime})\\
		+c^{2}\hat{\boldsymbol{\mathcal{B}}}^{\dagger}(\boldsymbol{x},\Omega)\boldsymbol{\otimes}\hat{\boldsymbol{\mathcal{B}}}^{\dagger}(\boldsymbol{x},\Omega^{\prime})
	\end{multline}
	We can interpret (\ref{stress-vacuum}) in a reasonably straightforward manner.  The final time--independent term is identical to the vacuum stress tensor used in Casimir--force calculations (see e.g.~\cite{volume9,philbin2011}).  This term is multiplied by \(\boldsymbol{R}-\boldsymbol{R}_{0}\) in (\ref{first-order-correction}), and consequently shifts the peak of the wave--packet in accordance with the Casimir force.  Clearly this force would be expected to vanish in the case of a lone body in free space.  Meanwhile, the first line is associated with the production of pairs of excitations propagating from the surface to within the volume of the dielectric body (c.f. (\ref{EBexp})), and is non--zero even for an isolated body in vacuum.
	\par
	As it stands, the term on the second line of (\ref{stress-vacuum}) is not finite.  This is because the integral over \(\Omega\) diverges, which is a well known feature of the vacuum stress tensor~\cite{volume9}.  Yet (\ref{first-order-correction}) contains the integral of \(\hat{\boldsymbol{\sigma}}|0\rangle\) over a closed surface, and the divergent contribution to (\ref{stress-vacuum}) often vanishes when integrated over a closed surface.  The standard procedure for dealing with the infinite contribution is to subtract a Green function for a homogeneous medium from the Green functions in (\ref{tau}).  This procedure is justified in appendix~\ref{appendix-B}, where it is also shown that the vacuum force on a lone object vanishes in free space.  Hence the second line of (\ref{stress-vacuum}) does not contribute to (\ref{first-order-correction}).
	\par
	In addition to \(\hat{\boldsymbol{\sigma}}\), (\ref{first-order-correction}) also contains the time derivative of \(\hat{\boldsymbol{\mathcal{P}}}_{A}|0\rangle\), which is,
	\begin{multline}
		\frac{\partial\hat{\boldsymbol{\mathcal{P}}}_{A}}{\partial t^{\prime}}|0\rangle=i\int_{0}^{\infty}d\Omega\int_{0}^{\infty}d\Omega^{\prime}e^{i(\Omega+\Omega^{\prime})t^{\prime}}(\Omega+\Omega^{\prime})\\
		\hat{\boldsymbol{\mathcal{E}}}^{\dagger}(\boldsymbol{x},\Omega)\boldsymbol{\times}\boldsymbol{\mathcal{H}}^{\dagger}(\boldsymbol{x},\Omega^{\prime})|0\rangle\label{ab-mom-op}
	\end{multline}
	where \(\boldsymbol{\mathcal{H}}^{\dagger}(\boldsymbol{x},\Omega^{\prime})=\mu_{0}^{-1}\hat{\boldsymbol{\mathcal{B}}}^{\dagger}(\boldsymbol{x},\Omega^{\prime})-\hat{\boldsymbol{\mathcal{M}}}^{\dagger}(\boldsymbol{x},\Omega^{\prime})\).  Combining (\ref{stress-vacuum}) and (\ref{ab-mom-op}) with (\ref{first-order-correction}) and performing the integration over time, we find,
	\begin{widetext}
		\begin{multline}
		|\chi(t)\rangle=\left(\frac{\sqrt{\alpha/\pi}}{1+i\hbar\alpha t/M}\right)^{3/2}e^{-\frac{\alpha (\boldsymbol{R}-\boldsymbol{R}_{0})^{2}}{2(1+i\hbar\alpha t/M)}}\bigg\{1-\frac{1}{2M}\int_{0}^{t}\left(\boldsymbol{\nabla}_{\boldsymbol{R}}\boldsymbol{\cdot}\hat{\boldsymbol{\Gamma}}\right)_{0}dt^{\prime}-(\boldsymbol{R}-\boldsymbol{R}_{0})\boldsymbol{\cdot}\int_{0}^{\infty} d\Omega\int_{0}^{\infty} d\Omega^{\prime}\frac{F(t,\Omega,\Omega^{\prime})}{\hbar(\Omega+\Omega^{\prime})}\\
		\times\left[\frac{i(\Omega+\Omega^{\prime})}{c^{2}}\int_{V}\hat{\boldsymbol{\mathcal{E}}}^{\dagger}(\boldsymbol{x},\Omega)\boldsymbol{\times}\hat{\boldsymbol{\mathcal{H}}}^{\dagger}(\boldsymbol{x},\Omega^{\prime})d^{3}\boldsymbol{x}-\epsilon_{0}\int_{\partial V}d\boldsymbol{S}\boldsymbol{\cdot}\left(\hat{\boldsymbol{\upsilon}}(\boldsymbol{x},\Omega,\Omega^{\prime})-\frac{1}{2}\boldsymbol{\mathbb{1}}_{3}\text{Tr}\left[\hat{\boldsymbol{\upsilon}}(\boldsymbol{x},\Omega,\Omega^{\prime})\right]\right)\right]\bigg\}|0\rangle
		\end{multline}
	\end{widetext}
	where,
	\begin{multline}
		F(t,\Omega,\Omega^{\prime})=e^{i(\Omega+\Omega^{\prime})t}\\
		-\frac{1}{1+i\hbar\alpha t/M}\left[1+\frac{\hbar\alpha}{M}\left(\frac{e^{i(\Omega+\Omega^{\prime})t}-1}{(\Omega+\Omega^{\prime})}\right)\right]\label{F-f}
	\end{multline}
	The wave-packet evolves into a sum of two amplitudes: the first for the field remaining in the vacuum state and the centre of mass spreading as would be predicted in the absence of any coupling to the field; and the second for pairs of field excitations to be created, and the centre of mass to be shifted by a quantity equal to the double frequency integral over the oscillatory factor, (\ref{F-f}) times a quadratic function of the expansion coefficients (see appendix~\ref{appendix-A}).  The magnitude of this fluctuation in the wave packet evidently depends on the size, shape, and dispersion of the object.   The relevant timescale of this fluctuation is one over the frequency where the imaginary part of the susceptibilities is largest. Typically this timescale is many orders of magnitude faster than the spreading time of the wave-packet, \(T_{s}\). 
%
%
	\section{Conclusions}
	It has been shown that the action (\ref{modified-action})---previously derived in~\cite{horsley2012}---can be used as the basis for both a classical and quantum theory of radiation pressure, fully including the effects of dispersion and dissipation.  The action is constructed so that it gives the Maxwell equations containing the correct constitutive relations for electromagnetism interacting with a moving medium.  Here we investigated varying this action with respect to the centre of mass coordinate to give equations of motion for the macroscopic body.
	\par
	We found that the expression for the classical force on a macroscopic body is given by (\ref{classical-result}).  In the case where the body is stationary this reduces to the integral of the Maxwell stress tensor over the surface of the body, minus the time derivative of the Abraham momentum within the body.  It was found that this is true even for dielectric bodies in contact, but that we must modify the action if we are to consider fluids or elastic media.
	\par
	Using the canonical procedure, a Hamiltonian was derived which was turned into a quantum mechanical operator and applied to the quantum motion of a small object.  We found that the average force on the body is equal to the expectation value of the operator analogue of the classical result (\ref{quantum-wave-packet-force}).  The detailed dynamics of an initially localised wave packet in vacuum were then analysed.  It was found that the peak of the wave packet is pushed by an amount equal to the usual expression for the Casimir force (in terms of electromagnetic Green functions), plus a time dependent factor corresponding to the creation of pairs of `polariton' excitations within the body.  The timescale of this fluctuation is determined by the frequency at which the object shows significant dissipation.
	\acknowledgements
	The author wishes to thank T. G. Philbin, M. Artoni, G. C. La Rocca, S. Kawka, M. Babiker, and S. Lloyd for many very useful discussions.  This work was financially supported by the EPSRC.
	\appendix
%
%
	\section{Normal mode expansion of field operators\label{appendix-A}}
	\par
	For reference here we give the expansion of the field operators that diagonalize (\ref{Hp}) for a fixed position of the centre of mass \(\boldsymbol{R}=\boldsymbol{R}_{0}\).  This expansion is used in section~\ref{wave-packet}, to diagonalize (\ref{Hp}), giving (\ref{H0}).  More details can be found on the derivation and verification of the validity of this expansion within~\cite{philbin2010,horsley2012}.  A set of bosonic operators are defined with the following commutation relations,
	\begin{equation}
		\left[\hat{\boldsymbol{C}}_{\omega}^{(\lambda)}(\boldsymbol{x}),\hat{\boldsymbol{C}}_{\omega^{\prime}}^{(\lambda^{\prime})\dagger}(\boldsymbol{x}^{\prime})\right]=\boldsymbol{\mathbb{1}}_{3}\delta_{\lambda\lambda^{\prime}}\delta^{(3)}(\boldsymbol{x}-\boldsymbol{x}^{\prime})\delta(\omega-\omega^{\prime})\label{bose-com}
	\end{equation}
	in order to define the number states associated with these operators, we write them in component form in terms of a set of unit vectors, \(\boldsymbol{e}_{\sigma}\)
	\begin{equation}
		\hat{\boldsymbol{C}}_{\omega}^{(\lambda)}(\boldsymbol{x})=\sum_{\sigma}\boldsymbol{e}_{\sigma}\hat{C}_{\omega,\sigma}^{(\lambda)}(\boldsymbol{x})\label{component-expansion}
	\end{equation}
	a set of states can then be defined in the usual way, with \(\hat{C}_{\omega,\sigma}^{(\lambda)}(\boldsymbol{x})|0\rangle=0\), \(\hat{C}_{\omega,\sigma}^{(\lambda)\dagger}(\boldsymbol{x})|0\rangle=|1_{\omega,\boldsymbol{x},\sigma,\lambda}\rangle\), etc.  The canonical variables in (\ref{Hp}) are expanded in terms of these bosonic operators.  The expansion takes the following form,
	\begin{equation}
		\hat{\boldsymbol{A}}(\boldsymbol{x},t)=\int_{0}^{\infty}d\Omega\left[\hat{\boldsymbol{\mathcal{A}}}(\boldsymbol{x},\Omega)e^{-i\Omega t}+\hat{\boldsymbol{\mathcal{A}}}^{\dagger}(\boldsymbol{x},\Omega)e^{i\Omega t}\right]\label{A-exp}
	\end{equation} 
	where, 
	\begin{multline}
		\hat{\boldsymbol{\mathcal{A}}}(\boldsymbol{x},\Omega)=\sum_{\lambda}\sqrt{\frac{\hbar}{2\Omega}}\int d^{3}\boldsymbol{x}^{\prime}\boldsymbol{f}_{\boldsymbol{A}}^{(\lambda)}(\boldsymbol{x},\boldsymbol{x}^{\prime},\Omega)\boldsymbol{\cdot}\hat{\boldsymbol{C}}_{\Omega}^{(\lambda)}(\boldsymbol{x}^{\prime})\label{A-exp-2}
	\end{multline}
	where the summation over \(\lambda\) runs over the two indices, \(E\) and \(B\) (see section~\ref{classical-section}).  In the case of the reservoir of oscillators, the notation is only slightly more complicated than (\ref{A-exp}--\ref{A-exp-2}), taking the form,
	\begin{multline}
		\hat{\boldsymbol{X}}_{\omega}^{(\lambda)}(\boldsymbol{x},t)=\int_{0}^{\infty}d\Omega\bigg[\hat{\boldsymbol{\mathcal{X}}}_{\omega}^{(\lambda)}(\boldsymbol{x},\Omega)e^{-i\Omega t}\\
		+\hat{\boldsymbol{\mathcal{X}}}^{(\lambda)\dagger}_{\omega}(\boldsymbol{x},\Omega)e^{i\Omega t}\bigg]
	\end{multline}
	where
	\begin{multline}
		\hat{\boldsymbol{\mathcal{X}}}_{\omega}^{(\lambda)}(\boldsymbol{x},\Omega)=\sum_{\lambda^{\prime}}\sqrt{\frac{\hbar}{2\Omega}}\int d^{3}\boldsymbol{x}^{\prime}\\
		\times\boldsymbol{f}_{\boldsymbol{X}}^{(\lambda\lambda^{\prime})}(\boldsymbol{x},\boldsymbol{x}^{\prime},\omega,\Omega)\boldsymbol{\cdot}\hat{\boldsymbol{C}}_{\omega}^{(\lambda^{\prime})}(\boldsymbol{x}^{\prime})\label{X-exp}\\
	\end{multline}
	In order that the change of variables (\ref{A-exp}--\ref{X-exp}) (along with the equivalent expressions for the canonical momentum) leave physical predictions unaltered, the canonical commutation relations, (\ref{em-com}--\ref{osc-com}) must be preserved.  One such expansion was found in~\cite{philbin2010}, which converts (\ref{Hp}) into (\ref{H0}).  In this case the expansion coefficients for the reservoir operators are given by,
	\begin{multline}
		\boldsymbol{f}_{\boldsymbol{X}}^{(\lambda\lambda^{\prime})}(\boldsymbol{x},\boldsymbol{x}^{\prime},\omega,\Omega)=\boldsymbol{\mathbb{1}}_{3}\delta_{\lambda\lambda^{\prime}}\delta(\omega-\Omega)\delta^{(3)}(\boldsymbol{x}-\boldsymbol{x}^{\prime})\\
		+\frac{\hat{\boldsymbol{O}}_{\lambda}(\boldsymbol{x},\omega)\boldsymbol{\cdot}\boldsymbol{G}(\boldsymbol{x},\boldsymbol{x}^{\prime};\Omega)\boldsymbol{\cdot}\hat{\boldsymbol{O}}^{\dagger}_{\lambda^{\prime}}(\boldsymbol{x}^{\prime},\Omega)}{(\omega-\Omega-i\eta)(\omega+\Omega+i\eta)}\label{fX}
	\end{multline}
	where \(\eta\) is an infinitesimal quantity that serves to impose retarded boundary conditions on the operators.  The canonical momenta operators of the reservoir are given by,
	\begin{equation}
		\boldsymbol{f}_{\boldsymbol{\Pi}_{\boldsymbol{X}}}^{(\lambda\lambda^{\prime})}(\boldsymbol{x},\boldsymbol{x}^{\prime},\omega,\Omega)=-i\Omega\boldsymbol{f}_{\boldsymbol{X}}^{(\lambda\lambda^{\prime})}(\boldsymbol{x},\boldsymbol{x}^{\prime},\omega,\Omega).\label{fPX}
	\end{equation}
	The two index quantity, \(\boldsymbol{G}\) in (\ref{fX}) is the electromagnetic Green function, and is a solution of,
	\begin{multline}
		\boldsymbol{\nabla}\boldsymbol{\times}\mu^{-1}(\boldsymbol{x},\Omega)\boldsymbol{\nabla}\boldsymbol{\times}\boldsymbol{G}(\boldsymbol{x},\boldsymbol{x}^{\prime};\Omega)-\Omega^{2}\epsilon(\boldsymbol{x},\Omega)\boldsymbol{G}(\boldsymbol{x},\boldsymbol{x}^{\prime};\Omega)\\
		=\boldsymbol{\mathbb{1}}_{3}\delta^{(3)}(\boldsymbol{x}-\boldsymbol{x}^{\prime})\label{green-eqn}
	\end{multline}
	where \(\epsilon=\epsilon_{0}+\chi_{\text{\tiny{EE}}}\), and \(\mu^{-1}=\mu_{0}^{-1}-\chi_{\text{\tiny{BB}}}\), with the susceptibilities defined as in section~\ref{classical-section}.  The operators, \(\hat{\boldsymbol{O}}_{\lambda}(\boldsymbol{x})\) are given by,
	\begin{align}
		\hat{\boldsymbol{O}}_{\text{\tiny{E}}}(\boldsymbol{x},\omega)&=i\Omega\alpha(\boldsymbol{x}-\boldsymbol{R}_{0},\omega)\boldsymbol{\mathbb{1}}_{3}\nonumber\\
		\hat{\boldsymbol{O}}_{\text{\tiny{B}}}(\boldsymbol{x},\omega)&=\beta(\boldsymbol{x}-\boldsymbol{R}_{0},\omega)\boldsymbol{\nabla}\boldsymbol{\times}\label{operators}
	\end{align}
	The curl operation given within \(\hat{\boldsymbol{O}}_{\text{\tiny{B}}}\) is such that it takes derivatives with respect to the coordinate given in the argument.  For example,
	\begin{multline}
		\hat{\boldsymbol{O}}_{\text{\tiny{E}}}(\boldsymbol{x},\omega)\boldsymbol{\cdot}\boldsymbol{G}(\boldsymbol{x},\boldsymbol{x}^{\prime};\Omega)\boldsymbol{\cdot}\hat{\boldsymbol{O}}^{\dagger}_{\text{\tiny{B}}}(\boldsymbol{x}^{\prime},\Omega)=\\
		i\Omega\alpha(\boldsymbol{x}-\boldsymbol{R}_{0},\omega)\beta(\boldsymbol{x}^{\prime}-\boldsymbol{R}_{0},\Omega)\boldsymbol{G}(\boldsymbol{x},\boldsymbol{x}^{\prime};\Omega)\boldsymbol{\times}\overleftarrow{\boldsymbol{\nabla}}^{\prime}
	\end{multline}
	From (\ref{operators}), the following identity can be derived which we give here for reference,
	\begin{multline}
		\sum_{\lambda}\hat{\boldsymbol{O}}_{\lambda}^{\dagger}(\boldsymbol{x},\Omega)\boldsymbol{\cdot}\hat{\boldsymbol{O}}_{\lambda}(\boldsymbol{x},\Omega)=\frac{2\Omega}{\pi}\bigg[\Omega^{2}\text{Im}[\chi_{\text{\tiny{EE}}}(\boldsymbol{x}-\boldsymbol{R}_{0},\Omega)]\boldsymbol{\mathbb{1}}_{3}\\
		+\boldsymbol{\times}\overleftarrow{\boldsymbol{\nabla}}\boldsymbol{\cdot}\text{Im}[\chi_{\text{\tiny{BB}}}(\boldsymbol{x}-\boldsymbol{R}_{0},\Omega)]\boldsymbol{\nabla}\boldsymbol{\times}\bigg]\label{opid}
	\end{multline}
	Finally, the canonical variables of the electromagnetic field are expanded as,
	\begin{equation}
		\boldsymbol{f}_{\boldsymbol{A}}^{(\lambda)}(\boldsymbol{x},\boldsymbol{x}^{\prime},\Omega)=\boldsymbol{G}_{\perp}(\boldsymbol{x},\boldsymbol{x}^{\prime};\Omega)\boldsymbol{\cdot}\hat{\boldsymbol{O}}_{\lambda}^{\dagger}(\boldsymbol{x}^{\prime},\Omega)\label{fA}
	\end{equation}
	and,
	\begin{multline}
		\boldsymbol{f}_{\boldsymbol{\Pi}_{\boldsymbol{A}}}^{(\lambda)}(\boldsymbol{x},\boldsymbol{x}^{\prime},\Omega)=-i\Omega\epsilon(\boldsymbol{x},\Omega)\boldsymbol{G}(\boldsymbol{x},\boldsymbol{x}^{\prime};\Omega)\boldsymbol{\cdot}\hat{\boldsymbol{O}}_{\lambda}^{\dagger}(\boldsymbol{x}^{\prime},\Omega)\\
		-\boldsymbol{\mathbb{1}}_{3}\alpha(\boldsymbol{x}-\boldsymbol{R}_{0})\delta_{\lambda\lambda^{\prime}}\delta^{(3)}(\boldsymbol{x}-\boldsymbol{x}^{\prime})
	\end{multline}
	The subscript, `\(\perp\)' in (\ref{fA}) implies taking the transverse part with respect to the first index, and the first argument, i.e. \(\boldsymbol{\nabla}\boldsymbol{\cdot}\boldsymbol{G}_{\perp}(\boldsymbol{x},\boldsymbol{x}^{\prime};\Omega)=0\).
	\par
	For reference we also give the expansion coefficients for the electric and magnetic field operators defined below (\ref{quantum-force}),
	\begin{align}
		\boldsymbol{f}_{\boldsymbol{E}}^{(\lambda)}(\boldsymbol{x},\boldsymbol{x}^{\prime},\Omega)&=i\Omega\boldsymbol{G}(\boldsymbol{x},\boldsymbol{x}^{\prime};\Omega)\boldsymbol{\cdot}\hat{\boldsymbol{O}}_{\lambda}^{\dagger}(\boldsymbol{x}^{\prime},\Omega)\nonumber\\
		\boldsymbol{f}_{\boldsymbol{B}}^{(\lambda)}(\boldsymbol{x},\boldsymbol{x}^{\prime},\Omega)&=\boldsymbol{\nabla}\boldsymbol{\times}\boldsymbol{G}(\boldsymbol{x},\boldsymbol{x}^{\prime};\Omega)\boldsymbol{\cdot}\hat{\boldsymbol{O}}_{\lambda}^{\dagger}(\boldsymbol{x}^{\prime},\Omega)\label{EBexp}
	\end{align}
	using (\ref{EBexp}) we then find that the commutation relations between the Fourier components of the electric and magnetic fields are,
	\begin{equation}
		\left[\hat{\boldsymbol{\mathcal{E}}}(\boldsymbol{x},\Omega),\hat{\boldsymbol{\mathcal{E}}}^{\dagger}(\boldsymbol{x}^{\prime},\Omega^{\prime})\right]=\delta(\Omega-\Omega^{\prime})\frac{\hbar\Omega^{2}}{\pi}\text{Im}[\boldsymbol{G}(\boldsymbol{x},\boldsymbol{x}^{\prime},\Omega)]\label{C1}
	\end{equation}
	and
	\begin{multline}
		\left[\hat{\boldsymbol{\mathcal{B}}}(\boldsymbol{x},\Omega),\hat{\boldsymbol{\mathcal{B}}}^{\dagger}(\boldsymbol{x}^{\prime},\Omega^{\prime})\right]=\\
		\delta(\Omega-\Omega^{\prime})\frac{\hbar}{\pi}\boldsymbol{\nabla}\boldsymbol{\times}\text{Im}[\boldsymbol{G}(\boldsymbol{x},\boldsymbol{x}^{\prime},\Omega)]\boldsymbol{\times}\overleftarrow{\boldsymbol{\nabla}}^{\prime}\label{C2}
	\end{multline}
	and finally,
	\[
		\left[\hat{\boldsymbol{\mathcal{E}}}(\boldsymbol{x},\Omega),\hat{\boldsymbol{\mathcal{B}}}^{\dagger}(\boldsymbol{x}^{\prime},\Omega^{\prime})\right]=\frac{i\hbar\Omega}{2}\delta(\Omega-\Omega^{\prime})\text{Im}[\boldsymbol{G}(\boldsymbol{x},\boldsymbol{x}^{\prime},\Omega)]\boldsymbol{\times}\overleftarrow{\boldsymbol{\nabla}}^{\prime}\label{C3}
	\]
%
%
	\section{Regularization, and the vacuum force on a lone object\label{appendix-B}}
	\par
	Here we illustrate the canonical method for making the vacuum stress tensor finite, and demonstrate that the vacuum force on an isolated body in free space is zero.  This vacuum stress tensor is given by
	\begin{multline}
		\boldsymbol{\sigma}_{0}=\frac{\hbar\epsilon_{0}}{\pi}\lim_{\boldsymbol{x}\to\boldsymbol{x}^{\prime}}\int_{0}^{\infty}d\Omega\,\text{Im}\bigg\{\boldsymbol{\tau}(\boldsymbol{x},\boldsymbol{x}^{\prime},\Omega)\\
		-\frac{1}{2}\boldsymbol{\mathbb{1}}_{3}\text{Tr}\left[\boldsymbol{\tau}(\boldsymbol{x},\boldsymbol{x}^{\prime},\Omega)\right]\bigg\}\label{s0}
	\end{multline}
	where \(\boldsymbol{\tau}(\boldsymbol{x},\boldsymbol{x}^{\prime},\Omega)\) is defined by (\ref{tau}).  As stated in the main text, (\ref{s0}) is not finite due to the behaviour of the integral over frequency.  To remedy this, one usually subtracts \(\boldsymbol{G}_{0}(\boldsymbol{x},\boldsymbol{x}^{\prime};\Omega)\) from the Green functions in (\ref{tau}),
	\begin{equation}
		\bar{\boldsymbol{G}}(\boldsymbol{x},\boldsymbol{x}^{\prime};\Omega)=\boldsymbol{G}(\boldsymbol{x},\boldsymbol{x}^{\prime};\Omega)-\boldsymbol{G}_{0}(\boldsymbol{x},\boldsymbol{x}^{\prime};\Omega)\label{regularized-G}
	\end{equation}
	For piecewise homogeneous media the quantity \(\boldsymbol{G}_{0}\) is the solution to (\ref{green-eqn}) for the case where \(\epsilon=\text{const.}\) and \(\mu=\text{const.}\)  The constant values of \(\epsilon\) and \(\mu\) are given by the local values where (\ref{s0}) is evaluated.  This modification cannot affect the value of \(\boldsymbol{\nabla}\boldsymbol{\cdot}\boldsymbol{\sigma}_{0}\), for \(\text{Im}[\boldsymbol{G}_{0}(\boldsymbol{x},\boldsymbol{x};\Omega)]\) does not depend on \(\boldsymbol{x}\) (it is defined for a homogeneous medium).
	\par
	We now take the imaginary part of (\ref{s0}) outside the integral over frequency, and look to rotate the integration from the positive real axis to the imaginary one.  This transformation is useful for the purpose of performing computations.  As \(|\Omega|\to\infty\) in the upper half complex plane, it can be shown from (\ref{green-eqn}) that \(\boldsymbol{G}\) tends to a quantity proportional to a delta function in position, divided by frequency squared.  This means that without applying (\ref{regularized-G}) to (\ref{s0}) both the integral over frequency and the limit of \(\boldsymbol{x}\to\boldsymbol{x}^{\prime}\) diverge everywhere.  However, subtracting \(\boldsymbol{G}_{0}\) from \(\boldsymbol{G}\) removes this divergent term, leaving one with, at worst terms proportional to \(\Omega^{-4}\) in \(\bar{\boldsymbol{G}}\).  Therefore, once (\ref{regularized-G}) has been applied, the integrand of (\ref{s0}) goes to zero as \(|\Omega|^{-2}\) or faster in the upper half plane.  This finding, along with the analyticity of the Green functions in the upper half plane mean we can thus rotate the integration from the positive real axis to the imaginary one, where the Green functions are real~\cite{volume8}.  Taking the imaginary part of the resulting expression we obtain,
	\begin{multline}
		\bar{\boldsymbol{\sigma}}_{0}=\frac{\hbar\epsilon_{0}}{\pi}\lim_{\boldsymbol{x}\to\boldsymbol{x}^{\prime}}\int_{0}^{\infty}d\xi\,\bigg[\bar{\boldsymbol{\tau}}(\boldsymbol{x},\boldsymbol{x}^{\prime},i\xi)\\
		-\frac{1}{2}\boldsymbol{\mathbb{1}}_{3}\text{Tr}\left[\bar{\boldsymbol{\tau}}(\boldsymbol{x},\boldsymbol{x}^{\prime},i\xi)\right]\bigg]\label{regularized-vacuum-stress}
	\end{multline}
	where the bars indicate that (\ref{regularized-G}) has been applied.  This is the expression commonly used in calculations of the Casimir force~\cite{volume9}.
	\par
	We now show that when sufficiently isolated from other bodies, the force on an object computed from (\ref{regularized-vacuum-stress}) vanishes.  First it is shown that the divergence of (\ref{regularized-vacuum-stress}) vanishes in free space.  To see this we take the divergence in the following manner before the limit \(\boldsymbol{x}\to\boldsymbol{x}^{\prime}\) is taken,
	\begin{multline}
		\boldsymbol{\nabla}\boldsymbol{\cdot}\bar{\boldsymbol{\sigma}}_{0}=\frac{\hbar\epsilon_{0}}{\pi}\lim_{\boldsymbol{x}\to\boldsymbol{x}^{\prime}}\int_{0}^{\infty}d\xi\,\bigg\{(\boldsymbol{\nabla}+\boldsymbol{\nabla}^{\prime})\boldsymbol{\cdot}\bar{\boldsymbol{\tau}}(\boldsymbol{x},\boldsymbol{x}^{\prime},i\xi)\\
		-\frac{1}{2}(\boldsymbol{\nabla}+\boldsymbol{\nabla}^{\prime})\text{Tr}\left[\bar{\boldsymbol{\tau}}(\boldsymbol{x},\boldsymbol{x}^{\prime},i\xi)\right]\bigg\}\label{div-s}
	\end{multline}
	To evaluate (\ref{div-s}) we note three things.  Firstly, in free space \(\bar{\boldsymbol{G}}\) satisfies (\ref{green-eqn}) with \(\mu=\mu_{0}\) and \(\epsilon=\epsilon_{0}\), and the right hand side set to zero.  Therefore \(\boldsymbol{\nabla}\boldsymbol{\cdot}\bar{\boldsymbol{G}}(\boldsymbol{x},\boldsymbol{x}^{\prime};i\xi)=0\). Secondly, the gradient of the trace of rank two tensor equals,
	\begin{align}
		\boldsymbol{\nabla}\text{Tr}[\bar{\boldsymbol{\tau}}]\equiv\partial_{i} \bar{\tau}_{jj}&=\epsilon_{i j k}\epsilon_{k l m}\partial_{l}\bar{\tau}_{j m}+\partial_{l}\bar{\tau}_{l i}\nonumber\\
		&\equiv\boldsymbol{\nabla}\boldsymbol{\cdot}\bar{\boldsymbol{\tau}}+2\left[\bar{\boldsymbol{\tau}}\boldsymbol{\times}\overleftarrow{\boldsymbol{\nabla}}\right]^{\star}\label{identity-1}
	\end{align}
	where repeated indices are summed over, and the star indicates that we have taken the dual of the two index quantity.  Finally, due to reciprocity, i.e. \(\boldsymbol{G}(\boldsymbol{x},\boldsymbol{x}^{\prime};\Omega)=\boldsymbol{G}^{T}(\boldsymbol{x}^{\prime},\boldsymbol{x};\Omega)\), we can interchange the two coordinates within the trace of \(\bar{\boldsymbol{\tau}}\), \(\boldsymbol{\nabla}\text{Tr}[\bar{\boldsymbol{\tau}}(\boldsymbol{x},\boldsymbol{x}^{\prime},i\xi)]=\boldsymbol{\nabla}\text{Tr}[\bar{\boldsymbol{\tau}}(\boldsymbol{x}^{\prime},\boldsymbol{x},i\xi)]\).  Combining these three pieces of information, and dropping a term that vanishes in the \(\boldsymbol{x}\to\boldsymbol{x}^{\prime}\) limit,
	\begin{multline}
		\boldsymbol{\nabla}\boldsymbol{\cdot}\bar{\boldsymbol{\sigma}}_{0}=-\frac{\hbar\epsilon_{0}}{\pi}\lim_{\boldsymbol{x}\to\boldsymbol{x}^{\prime}}\int_{0}^{\infty}d\xi\,\bigg\{\left[\bar{\boldsymbol{\tau}}(\boldsymbol{x}^{\prime},\boldsymbol{x},i\xi)\boldsymbol{\times}\overleftarrow{\boldsymbol{\nabla}}^{\prime}\right]^{\star}\\
		+\left[\bar{\boldsymbol{\tau}}(\boldsymbol{x},\boldsymbol{x}^{\prime},i\xi)\boldsymbol{\times}\overleftarrow{\boldsymbol{\nabla}}\right]^{\star}\bigg\}\label{div-stress}
	\end{multline}
	Inserting the explicit form for \(\bar{\boldsymbol{\tau}}\) and applying (\ref{green-eqn}) we can evaluate the dual quantities in (\ref{div-stress}),
	\begin{multline}
		\left[\bar{\boldsymbol{\tau}}(\boldsymbol{x},\boldsymbol{x}^{\prime},i\xi)\boldsymbol{\times}\overleftarrow{\boldsymbol{\nabla}}^{\prime}\right]^{\star}=\frac{\Omega^{2}}{2}\bigg\{(\boldsymbol{\nabla}-\boldsymbol{\nabla}^{\prime})\text{Tr}[\bar{\boldsymbol{G}}(\boldsymbol{x},\boldsymbol{x}^{\prime};i\xi)]\\
		+\boldsymbol{\nabla}\boldsymbol{\cdot}\bar{\boldsymbol{G}}(\boldsymbol{x}^{\prime},\boldsymbol{x};i\xi)-\boldsymbol{\nabla}^{\prime}\boldsymbol{\cdot}\bar{\boldsymbol{G}}(\boldsymbol{x},\boldsymbol{x}^{\prime};i\xi)\bigg\}\label{identity-2}
	\end{multline}
	Inserting (\ref{identity-2}) into (\ref{div-stress}) and taking the limit \(\boldsymbol{x}\to\boldsymbol{x}^{\prime}\), we find the result that, in free space,
	\begin{equation}
		\boldsymbol{\nabla}\boldsymbol{\cdot}\bar{\boldsymbol{\sigma}}_{0}=0\label{div-zero}
	\end{equation}
	It was found in the main text that after inserting (\ref{stress-vacuum}) into (\ref{first-order-correction}) there is a term in the wave--function equal to the integral of \(\boldsymbol{\nabla}\boldsymbol{\cdot}\bar{\boldsymbol{\sigma}}_{0}\) over the dielectric body of interest.  If the body is alone in free space (or at least very far from neighbouring bodies) then due to (\ref{div-zero}) we can extend the integration domain over a very large region of space without altering the value of the integral.  Therefore the value of the integral of \(\boldsymbol{\nabla}\boldsymbol{\cdot}\bar{\boldsymbol{\sigma}}_{0}\) over the body is equal to \(\int d\boldsymbol{S}\boldsymbol{\cdot}\bar{\boldsymbol{\sigma}}_{0}\) over the surface of a sphere of large radius, centred on the body.  To conclude we show that this vanishes.
	\par
	Writing \(\boldsymbol{G}=\bar{\boldsymbol{G}}+\boldsymbol{G}_{0}\) and inserting this into (\ref{green-eqn}), we find that \(\bar{\boldsymbol{G}}\) obeys the differential equation
	\begin{multline}
		\boldsymbol{\nabla}\boldsymbol{\times}\mu^{-1}(\boldsymbol{x},i\xi)\boldsymbol{\nabla}\boldsymbol{\times}\bar{\boldsymbol{G}}(\boldsymbol{x},\boldsymbol{x}^{\prime};i\xi)+\xi^{2}\epsilon(\boldsymbol{x},i\xi)\bar{\boldsymbol{G}}(\boldsymbol{x},\boldsymbol{x}^{\prime};i\xi)\\
		=\boldsymbol{\nabla}\boldsymbol{\times}\chi_{\text{\tiny{BB}}}(\boldsymbol{x},i\xi)\boldsymbol{\nabla}\boldsymbol{\times}\boldsymbol{G}_{0}(\boldsymbol{x},\boldsymbol{x}^{\prime};i\xi)\\
		-\xi^{2}\chi_{\text{\tiny{EE}}}(\boldsymbol{x},i\xi)\boldsymbol{G}_{0}(\boldsymbol{x},\boldsymbol{x}^{\prime};i\xi)
	\end{multline}
	which has the solution
	\begin{widetext}
		\begin{equation}
			\bar{\boldsymbol{G}}(\boldsymbol{x},\boldsymbol{x}^{\prime};i\xi)=\int_{V} d^{3}\boldsymbol{x}^{\prime\prime}\boldsymbol{G}(\boldsymbol{x},\boldsymbol{x}^{\prime\prime};i\xi)\boldsymbol{\cdot}\left\{\boldsymbol{\times}\overleftarrow{\boldsymbol{\nabla}}^{\prime\prime}\boldsymbol{\cdot}\chi_{\text{\tiny{BB}}}(i\xi)\boldsymbol{\nabla}^{\prime\prime}\boldsymbol{\times}-\xi^{2}\chi_{\text{\tiny{EE}}}(i\xi)\boldsymbol{\mathbb{1}}_{3}\right\}\boldsymbol{\cdot}\boldsymbol{G}_{0}(\boldsymbol{x}^{\prime\prime},\boldsymbol{x}^{\prime};i\xi)
		\end{equation}
	\end{widetext}
	where \(\boldsymbol{G}_{0}(\boldsymbol{x},\boldsymbol{x}^{\prime};\Omega)\) is the free space Green function.  To leading order both \(\boldsymbol{G}(\boldsymbol{x},\boldsymbol{x}^{\prime\prime};\Omega)\) and \(\boldsymbol{G}_{0}(\boldsymbol{x},\boldsymbol{x}^{\prime\prime};\Omega)\) go to zero as \(|\boldsymbol{x}-\boldsymbol{x}^{\prime\prime}|^{-1}\exp{(-\xi|\boldsymbol{x}-\boldsymbol{x}^{\prime\prime}|)}\) at large distances.  After the integral over \(\xi\) in (\ref{regularized-vacuum-stress}) has been performed one has a quantity going to zero faster than \(|\boldsymbol{x}-\boldsymbol{x}^{\prime\prime}|^{-2}\), and \(\int d\boldsymbol{S}\boldsymbol{\cdot}\bar{\boldsymbol{\sigma}}_{0}\) therefore vanishes as the integration surface is taken to infinity.  We thus conclude that, as anticipated, the integral of \(\boldsymbol{\nabla}\boldsymbol{\cdot}\boldsymbol{\sigma}_{0}\) over the volume of an isolated body is zero.
%
%
	\bibliography{refs}
	\end{document}